\documentclass[twocolumn,english]{revtex4-2}
\usepackage[T1]{fontenc}
\usepackage[latin9]{inputenc}
\usepackage{color}
\usepackage{amsmath}
\usepackage{graphicx}
\usepackage{comment}
\usepackage{graphicx}
\usepackage{sidecap}
\usepackage[colorlinks=true, allcolors=red]{hyperref}

\makeatletter
\usepackage{babel}

\makeatother

\usepackage{babel}
\begin{document}
\title{ Finite Ion Temperature Effects on the Merging of Current-Carrying ELM Filaments in the edge region of a tokamak}

\author{Souvik Mondal$^{1,3,a}$}
\email{$^{a)}$Email: souvik.mondal@ipr.res.in} 
\author{N Bisai$^{1,3}$}
\author{Abhijit Sen$^{1,3}$}
\author{Indranil Bandyopadhyay$^{1,2,3}$}

\affiliation{$^1$Institute for Plasma Research, Bhat, Gandhinagar 382428, Gujarat, India \\
$^2$ITER-India, Institute for Plasma Research, Bhat, Gandhinagar 382428, Gujarat, India \\
$^3$Homi Bhabha National Institute, Training School Complex, Anushaktinagar, Mumbai 40094, India}

\begin{abstract}
Edge-localized-mode (ELM) filaments are crucial for cross-field transport at the tokamak edge; yet, their dynamics are often analyzed using the cold-ion approximation, despite experimental data indicating that $T_i \sim T_e$. This study employs a normalized three-dimensional fluid model to investigate the influence of finite ion temperature on the dynamics of unidirectional current-carrying ELM-like filaments. We demonstrate that increasing ion temperature substantially alters filament propagation and interaction, resulting in a delay of filament merging despite an increase in total kinetic energy due to a stronger pressure-gradient drive. The examination of single-filament dynamics indicates that finite ion temperature generates asymmetric potential structures, strong poloidal flows, and persistent rotational motion, which channel kinetic energy from radial propagation into vortical dynamics. A comprehensive examination of the ion-to-electron temperature ratio reveals a distinct transition from radially dominated to rotation-dominated behavior as ion temperature increases. These results provide a unified physical explanation for reduced radial transport and delayed merging in the warm-ion domain, emphasizing the necessity of incorporating ion temperature effects in the modeling of ELM filament dynamics and edge plasma transport.
\end{abstract}

\maketitle

\section{INTRODUCTION}

Edge-localized modes (ELMs) and the associated filamentary structures play a central role in cross-field transport in the edge and scrape-off-layer (SOL) regions of magnetically confined plasmas. These filaments, often referred to as ``blobs'' \cite{sarazin_intermittent_1998, Umansky_1998, KRASHENINNIKOV2001368, Zweben_2002, J_L_Terry_2003, Bisai_2004, bisai_3f2f_2004, Bisai_2005} or ELM filaments \cite{H_Zohm_1996_ELM, Wang_2013, Kass_1998, DIIID_ELM_PRL, RFX_ELM_PRL}, carry enhanced density, pressure, and parallel current, and are responsible for intermittent particle and heat fluxes to plasma-facing components. Understanding the dynamics of these filaments, including their propagation, interaction, and merging, is therefore essential for predictive modeling of edge transport and for the development of mitigation strategies in future fusion devices.

ELMs are typically categorized into many categories based on their dimensions, frequency, and fundamental instability processes. Type-I ELMs are large, low-frequency occurrences induced by peeling-ballooning instabilities, whereas Type-II and Type-III ELMs are defined by reduced amplitude but significantly higher frequency \cite{H_Zohm_1996_ELM}. In Type-II and Type-III regimes, the edge plasma demonstrates a quasi-continuous emission of filamentary structures, resulting in a larger accumulation of filaments at the edge. Consequently, interactions among filaments, such as collisions, clustering, and merging, are anticipated to occur with greater frequency and to significantly influence total transport dynamics.

Early theoretical and numerical studies of blob dynamics established the interchange instability as the primary mechanism driving radial filament propagation through $E\times B$ motion \cite{Garcia2006,dippolito_convective_2011}. Subsequent work demonstrated that filament behavior is strongly influenced by sheath dissipation, parallel current closure, and electromagnetic effects, leading to a wide variety of propagation regimes \cite{KRASHENINNIKOV_2008}. More recently, attention has turned to the interaction between multiple filaments, motivated by experimental observations of filament clustering and coalescence. In particular, numerical studies in the cold-ion limit have shown that unidirectional current-carrying filaments can attract each other and merge efficiently through direct radial convergence driven by interchange dynamics and current-current interaction \cite{myra_current_carrying_filament_2007,souvik_pop}.

At the same time, a growing body of experimental and theoretical work has demonstrated that finite ion temperature plays a crucial role in filament dynamics. Experimental studies have reported clear signatures of a warm-ion regime in the SOL, with ion temperatures comparable to or exceeding electron temperatures \cite{Kocan2012,Manz2013,Manz2015}. These observations motivated the development of warm-ion blob models, which showed that ion temperature and finite-Larmor-radius effects can substantially modify filament propagation. Gyrofluid and gyrokinetic simulations demonstrated that warm-ion filaments exhibit reduced radial velocity, enhanced poloidal motion, and the emergence of coherent rotational and vortical structures \cite{Jovanovic2008,Madsen2011_gyrofulid,Wiesenberger2014,Held2016,Ricci2013_gyrokinetic}. Scaling laws incorporating ion temperature effects have been proposed and successfully compared with experiments, highlighting the role of pressure-gradient-driven polarization currents in governing filament motion in the warm-ion regime \cite{Manz2013,Manz2015}.

Despite this progress, current warm-ion research has predominantly concentrated on the dynamics of isolated filaments. The impact of finite ion temperature on filament-filament interaction and merger is mostly unexplored. The gap is particularly for ELM filaments in the Type-II and Type-III regimes, where the high generation rate of filaments results in frequent interaction and collective phenomena. Furthermore, these filaments are distinguished by significant pressure fluctuations as well as high unidirectional parallel currents. Recent experimental and modeling suggest that current-carrying filaments develop in conditions when warm-ion effects are unavoidable \cite{Kocan2012,Held2016,Ahmed2025}. Nonetheless, the majority of computational studies on filament merging thus far have been confined to the cold-ion limit, leaving unresolved the impact of ion temperature on the interaction geometry, energy transfer, and coalescence of current-carrying filaments.

In this work, we address this problem by systematically investigating the influence of finite ion temperature on the dynamics and merging of unidirectional current-carrying ELM-like filaments using a normalized three-dimensional fluid model. We first compare the merging behavior of two filaments in cold- and warm-ion regimes and demonstrate that finite ion temperature leads to a delay in merging despite an increase in total kinetic energy. We also analyze the dynamics of a single isolated filament and show that warm-ion effects generate asymmetric potential structures, strong radial electric fields, and sustained rotational motion, resulting in a redistribution of kinetic energy from radial propagation into poloidal and vortical flows. Finally, by performing a systematic scan of the ion-to-electron temperature ratio ($\tau$), we establish a robust transition from radially dominated to rotation-dominated filament dynamics with increasing ion temperature.


This paper is organized as follows. Section~II describes the normalized three-dimensional fluid model and provides a qualitative discussion of the dynamics of ELM-like, current-carrying filaments. Section~III outlines the numerical setup, including the simulation domain, input parameters, initial conditions, and boundary conditions. In Section~IV, we present and discuss the simulation results, focusing on the merging of current-carrying filaments, the role of finite ion temperature in single-filament dynamics, and the systematic dependence of filament behavior on the ion-to-electron temperature ratio. Finally, Section~V summarizes the main findings and discusses their implications for ELM filament dynamics and edge plasma transport.

\section{MODEL EQUATIONS\label{sec:Model_Equations}}

To investigate the dynamics of filamentary ELM blobs, we employ a reduced fluid description derived from the Braginskii equations~\cite{braginskii1965transport}. A three-dimensional, two-fluid framework is used, formulated in Cartesian coordinates with the magnetic field directed along the $z$-axis, while the $x$ and $y$ directions represent the radial and poloidal directions, respectively.  

Unlike earlier cold-ion treatments where $T_i \rightarrow 0$, here we explicitly retain a finite ion temperature ($T_i \neq 0$). This is essential because in many experimental conditions $T_i \sim T_e$ in the edge region, and ion dynamics strongly modify blob propagation. In our model, both electrons and ions are treated as isothermal but with finite $T_e$ and $T_i$, so that $\tau_i = T_i/T_e$ enters naturally in the closure relations.  

In experiments, $T_e$ is not uniform along and across magnetic field lines, producing radial electric fields and $E \times B$ drifts of plasma filaments. In this work, however, we are primarily concerned with the nonlinear evolution of ELM blobs under the influence of ion temperature and parallel current, so we simplify the treatment by focusing on the dominant terms. We also allow for finite plasma beta, $\beta = 8\pi n T/B^2$, such that the inductive component of the parallel electric field,
\begin{equation}
    E_{\parallel}^{ind} = - \frac{1}{c}\frac{\partial A_{\parallel}}{\partial t},
\end{equation}
is retained, thereby including electromagnetic effects.  

The governing equations follow from particle conservation, current continuity (quasi-neutrality), the parallel electron momentum balance, and Maxwell's equations. Under these approximations the reduced model becomes \cite{lee_electromagnetic_2015, stepanenko_impact_2020}:
\begin{equation}
	\frac{e\rho_{s}^{2}}{T_{e}}n\frac{d\omega_i}{dt}
	= \frac{1}{e}\nabla_{\|}J_{\|}-(1+\tau)\frac{g_i}{\Omega_{s}}\frac{\partial n}{\partial y}, 
	\label{eq:unnormalized_vorticity}
\end{equation}
\begin{equation}
	\frac{dn}{dt}
	= \frac{1}{e}\nabla_{\|}J_{\|}-\frac{g_i}{\Omega_{s}}\frac{\partial n}{\partial y},
	\label{eq:unnormalized_continuity}
\end{equation}
\begin{equation}
	-\frac{e}{m_e c}\frac{dA_{\|}}{dt}
	= \frac{e}{m_{e}}\frac{\partial\phi}{\partial z}
	- \frac{T_{e}}{m_{e}}\nabla_{\|}\ln n
	+ \frac{e}{\sigma_{\|}m_{e}}J_{\|}.
	\label{eq:vector_potential}
\end{equation}

Equations~(\ref{eq:unnormalized_vorticity})-(\ref{eq:vector_potential}) originate from current continuity, density conservation, and generalized Ohm's law. Here $n$ is the plasma density, $\phi$ the electrostatic potential, $\omega = \nabla_{\perp}^2\phi+ (1/n_ee)\nabla_{\perp}^2p_i$ the vorticity, and $A_{\|}$ the parallel component of the magnetic vector potential. The characteristic scales are $\rho_s = c_s/\Omega_s$, with $c_s = \sqrt{T_e/m_i}$ the sound speed and $\Omega_s$ the ion gyro-frequency. The effective curvature drive is given by $g_i = 2c_s^2/R$, with $R$ the major radius of the tokamak.  

The parallel current density is defined as $J_{\|} = ne(V_{i\|}-V_{e\|})$, while the parallel electrical conductivity is $\sigma_{\|} = 1.96n_{0}e^{2}/m_{e}\nu_{ei}$, where the collision frequency is $\nu_{ei}=2.9\times10^{-6}n_{0}\ln\Lambda/T_{e}^{3/2}$ (with Coulomb logarithm $\ln\Lambda \approx 10$). The total derivative operator is
\begin{equation}
    \frac{d}{dt} = \frac{\partial}{\partial t} 
    + \frac{c}{B}\hat{b}_0 \times \nabla \phi \cdot \nabla,
\end{equation}
and the parallel gradient operator is
\begin{equation}
    \nabla_{\parallel} = \frac{\partial}{\partial z}
    + \left(\frac{\nabla A_{\parallel}}{B_0}\right)\times \hat{b}_0 \cdot \nabla,
\end{equation}
where $\hat{b}_0$ is the unit vector along the background magnetic field. Maxwell's equation gives the relation between $J_\parallel$ and $A_\parallel$:
\begin{equation}
J_\parallel = -\frac{c}{4\pi}\nabla_{\perp}^2A_\parallel.
\end{equation}

In the macroscopic blob regime, the perpendicular advection time $\tau_{\perp}\sim \delta/v_{E\times B}$ is typically much longer than the electron-ion collision time $\tau_{ei}$, justifying the neglect of electron inertia~\cite{stepanenko_impact_2020}. For the two-dimensional reduced model employed here, averaging along the field line ($\nabla_{\|}\sim 1/L_{\|}$) is performed, while retaining the inductive bracket term from $\nabla_{\|}A_{\|}$. This yields the approximate closure:
\begin{equation}
\nabla_{\|}J_{\|} = \frac{1}{B_0}[J_{\|}, A_{\|}]_{x,y},
\qquad
\nabla_{\|}\ln n = \frac{1}{B_0}[\ln n, A_{\|}]_{x,y}.
\end{equation}

To numerically simulate 
Eqs.~(\ref{eq:unnormalized_vorticity})-(\ref{eq:vector_potential}), we use the following normalizations:
${n}/{n_{0}}=\hat{n}$, $t\Omega_{s}=\hat{t}$, ${(x,y)}/{\rho_{s}}=(\hat{x},\hat{y})$, $v_{\parallel}/{c_{s}}=\hat{v}_{\parallel}$, ${J_{\parallel}}/{n_{0}ec_{s}}=\hat{J}_{\parallel}$, ${e\phi}/{T_{e0}}=\hat{\phi}$, ${A}/{B_{0}\rho_{s}}=\hat{A}$, ${T_{e}}/{T_{e0}}=\hat{T}_e$ and $\rho_s{\nabla_{\parallel}}=\hat{\nabla}_{\parallel}$ where $n_0$, $T_{e0}$, and $B_0$ represent the values of plasma density, electron temperature, and toroidal magnetic field in the Last Closed Flux Surface (LCFS). $v_\|$ represents the parallel velocity of the electrons.\\

Using the above approximations and normalizations, the final form of the equations can be expressed as (omitting hat):
\begin{equation}
\frac{\partial n}{\partial t}=-[\phi,n]+\frac{\partial J_{\|}}{\partial z}-\left[A_{\|},J_{\|}\right]-g\frac{\partial n}{\partial y},\label{eq:norm_density_continuty_eq}
\end{equation}
\begin{equation}
n\frac{\partial \omega}{\partial t}=-[\phi,\nabla_{\perp}^{2}\phi]+\frac{\partial J_{\|}}{\partial z}-\left[A_{\|},J_{\|}\right]-(1+\tau)g\frac{\partial n}{\partial y},\label{eq:norm_vorticity_eq}
\end{equation}
\begin{equation}
\frac{\partial A_{\|}}{\partial t}=\frac{\partial}{\partial z}(\ln n-\phi)-\left[A_{\|},\ln n\right]+\eta \nabla_{\perp}^{2}A_{\|},\label{eq:norm_vector_pot_eq}
\end{equation}
and 
\begin{equation}
J_\parallel=-a\nabla_{\perp}^2A_\parallel,\label{eq:norm_maxwell_eq}
\end{equation}
where $\omega=\nabla_{\perp}^{2}\phi + \tau(\nabla_{\perp}^{2}n)/n $ is the normalized vorticity, $g=\rho_s /R$ is the normalized effective gravity obtained form the normalization of $g_i$, $\eta=1/(\Omega_s \tau_s)$ is the normalized magnetic diffusion constant, $\tau_{s}=4\pi\sigma_{\|}\rho_{s}^{2}/c^{2}$ is the magnetic screen time, and $a=  1.96{m_{i}}/{m_{e}\nu_{ie}\tau_{s}}$.


\section{Numerical Simulation}
\label{sec:Numerical_simulation}


The numerical simulations of Eqs.~(\ref{eq:norm_density_continuty_eq})-(\ref{eq:norm_maxwell_eq}) are performed using typical edge plasma parameters relevant to high-$\beta$ tokamak conditions, as summarized in Table~\ref{table:1}. The chosen parameters correspond to plasma conditions near the last closed flux surface (LCFS) and are consistent with previous electromagnetic blob and filament studies \cite{lee_electromagnetic_pop}. All quantities are expressed in normalized units unless stated otherwise.

\begin{table}
\centering
\setlength{\tabcolsep}{12pt}
\begin{tabular}{l c c} 
 \hline
 Parameter & Value & Unit \\ [0.5ex] 
 \hline\hline
 $n_0$ & $1\times10^{14}$ & cm$^{-3}$ \\ 
 $T_e$ & 200 & eV \\
 $B_0$ & $5.3\times 10^4$ & G \\
 $R$ & 600 & cm \\
 $L$ & $10^4$ & cm \\
 $c_s$ & $9.98\times10^{6}$ & cm/s \\ 
 $\Omega_s$ & $2.64\times10^{8}$ & s$^{-1}$ \\
 $\rho_s$ & $3.78\times10^{-2}$ & cm \\
 $\nu_{ei}$ & $1.02\times10^{6}$ & s$^{-1}$ \\
 $\sigma_{\|}$ & $4.82\times10^{16}$ & s$^{-1}$ \\
 $g_i$ & $3.3\times 10^{11}$ & cm/s$^{2}$\\
 $g$ & $1.25\times 10^{-4}$ & normalized \\
 $\delta$ & 0.7 & cm \\ [1ex] 
 \hline
\end{tabular}
\caption{Typical high-$\beta$ plasma parameters relevant to ITER-like edge conditions. Plasma parameters near the LCFS are used in the simulations \cite{lee_electromagnetic_pop}.}
\label{table:1}
\end{table}

To investigate both filament-filament interaction and the underlying single-filament dynamics, we perform two sets of simulations: (i) double-blob simulations, where two filaments are initialized at different poloidal locations, and (ii) single-blob simulations, where only one filament is present. The simulation domain lengths are $L_x=L_y=256\rho_s$ and $L_z=256384\rho_s$ in the radial, poloidal, and toroidal directions, respectively.

\subsubsection*{A. Double-blob initial condition}

For the double-blob simulations, two filaments are initialized with Gaussian density perturbations in the perpendicular plane at $t=0$. The initial density profile is given by
\begin{equation}
\begin{split}
n(\textbf{r},0) = 1 + n_b \exp\left[-\frac{(x-x_0)^2}{\delta^2}\right]\cdot 
\left[\exp\left(-\frac{(y-y_1)^2}{\delta^2}\right) \right.\\
\left. + \exp\left(-\frac{(y-y_2)^2}{\delta^2}\right)\right]
\end{split}
\end{equation}
where $x_0=L_x/2$ is the initial radial position of the filament centers, and the poloidal locations are chosen as $y_1=0.325L_y$ and $y_2=0.675L_y$. The parameter $n_b$ denotes the relative amplitude of the density perturbation and is set to $n_b=2$, consistent with experimentally observed blob amplitudes.

The characteristic cross-sectional size of the filaments is denoted by $\delta$. In the present simulations, $\delta$ is chosen as
\[
\delta = \rho_s \left(\frac{g_i L^2}{4c_s^2 \rho_s}\right)^{1/5},
\]
which corresponds to the most stable characteristic blob size. This yields $\delta \approx 0.7$~cm, in agreement with typical blob sizes observed in tokamak experiments \cite{dippolito_convective_2011}.

The corresponding equilibrium parallel current density is initialized with the same perpendicular spatial structure as the density perturbation \cite{myra_current_carrying_filament_2007},
\begin{equation}
\begin{split}
J_\|(\textbf{r},0) = J_0 \exp\left[-\frac{(x-x_0)^2}{\delta^2}\right]\cdot 
\left[\exp\left(-\frac{(y-y_1)^2}{\delta^2}\right) \right.\\
\left. + \exp\left(-\frac{(y-y_2)^2}{\delta^2}\right)\right]
\end{split}
\end{equation}
where $J_0$ is the amplitude of the equilibrium parallel current density. This setup allows us to study the electromagnetic interaction and merging dynamics of unidirectional current-carrying filaments.

\subsubsection*{B. Single-blob initial condition}

For the single-blob simulations, a single filament is initialized at $t=0$ with a Gaussian density perturbation centered at $(x_0,y_0)$ in the perpendicular plane. The initial density profile is given by
\begin{equation}
n(\mathbf{r},0) = 1 + n_b \exp\left[-\frac{(x-x_0)^2+(y-y_0)^2}{\delta^2}\right],
\end{equation}
with $x_0=L_x/2$ and $y_0=L_y/2$. The parameters $n_b$ and $\delta$ are chosen to be identical to those used in the double-blob simulations, ensuring a consistent comparison between single-filament and interacting-filament dynamics. The equilibrium parallel current density for the single-blob case is initialized as
\begin{equation}
J_{\|}(\mathbf{r},0) = J_0 \exp\left[-\frac{(x-x_0)^2+(y-y_0)^2}{\delta^2}\right].
\end{equation}
These simulations are used to isolate the effects of finite ion temperature on filament structure, velocity, electric fields, and energetic partitioning without the influence of filament-filament interaction.

In both single- and double-blob simulations, the equilibrium parallel magnetic vector potential $A_{\|}$ is obtained numerically from Eq.~(\ref{eq:norm_maxwell_eq}) using Laplace inversion at the initial time. The initial conditions for the electrostatic potential $\phi$ and vorticity $\nabla_\perp^2\phi$ are set to zero. Periodic boundary conditions are applied in the poloidal ($y$) direction, while Neumann boundary conditions are imposed in the radial ($x$) and toroidal ($z$) directions for all evolved fields, including $n$, $\phi$, $\nabla_\perp^2\phi$, $J_{\|}$, and $A_{\|}$.

The model equations are solved numerically using the BOUT++ framework \cite{DUDSON_bout++}. Spatial derivatives are computed using fourth-order central difference schemes in the $x$ and $z$ directions, while Fourier-based methods are employed in the $y$ direction. Upwind derivatives are evaluated using a third-order weighted essentially non-oscillatory (WENO) scheme. Time integration is performed using the CVODE solver. The typical grid resolution used in the simulations is $N_x \times N_y \times N_z = 516 \times 512 \times 32$, which ensures good conservation of total energy. The corresponding grid spacings are $dx = dy = 0.5\rho_s$ and $dz = 8012\rho_s$.


\section{Simulation results}


In this section, we present numerical simulations of ELM-like current-carrying filaments using the normalized three-fluid model described in Sec.~II. The primary objective is to investigate how finite ion temperature modifies the dynamics of isolated filaments and their subsequent interaction and merging behavior. Particular emphasis is placed on the redistribution of kinetic energy between radial and poloidal motion, the emergence of rotational dynamics, and the resulting impact on radial convergence and filament merging. The cold-ion limit is used as a reference case, and the results are systematically extended to warm-ion regimes through a scan in the ion-to-electron temperature ratio $\tau=T_i/T_e$. Where relevant, the observed trends are discussed in the context of existing warm-ion blob scaling studies~\cite{Manz2013}, while highlighting the distinct role played by unidirectional parallel currents characteristic of ELM filaments.


\subsection{Modification of Filament Merging by Finite Ion Temperature Effects}


\begin{figure*}
    \centering
    \includegraphics[width=0.9\linewidth]{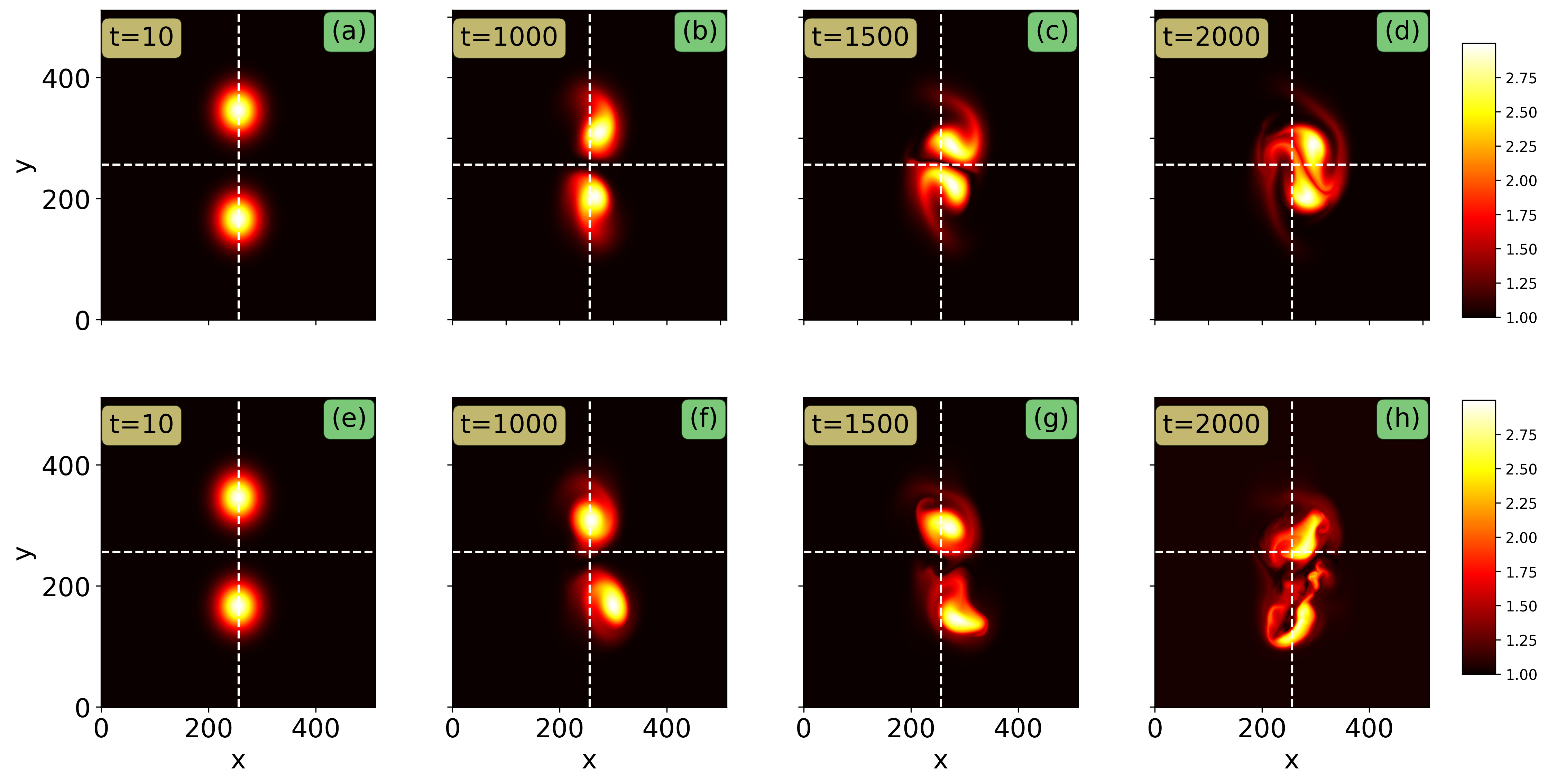}
    \caption{Time evolution of the density during the merging of two unidirectional current-carrying filaments for (a-d) the cold-ion case ($\tau=0$) and (e-h) the warm-ion case ($\tau=1.0$). Finite ion temperature leads to strong deformation and rotational motion of the density structures, delaying their direct merging.}
    \label{fig:density_merging}
\end{figure*}

Figure~\ref{fig:density_merging} shows the time evolution of the density for two unidirectional current-carrying filaments in the cold-ion ($\tau=0$) and warm-ion ($\tau=1.0$) regimes. At $t=0$, both cases consist of two identical, well-localized density perturbations symmetrically positioned about the mid-plane with the same initial separation, having a magnitude of initial current density is 0.75 MA/m$^2$. In the cold-ion case (a-d), the density structures remain compact and largely symmetric as they evolve. The filaments propagate predominantly in the poloidal direction and comes closer to each other directly, leading to a rapid overlap of the density peaks and the formation of a single merged structure. The absence of significant deformation or tilting of the density contours indicates that the interaction is dominated by radial interchange-driven motion with minimal rotational influence.

Conversely, the warm-ion scenario (e-h) demonstrates significant deformation of the density structures throughout the interaction. Over time, the filaments undergo elongation and tilting, and change their shapes considerably from the initial bi-Gaussian profiles. The density contours clearly show shear and rotation, with the filaments partially orbiting each other instead of directly merging. This behavior suggests that $\tau$  introduces an additional dynamic factor in the poloidal direction, allowing density perturbations to maintain rotational motion. As a result, the merging of the density filament is delayed, indicating that filament merging is less effective in the warm-ion regime.

\begin{figure*}
    \centering
    \includegraphics[width=0.9\linewidth]{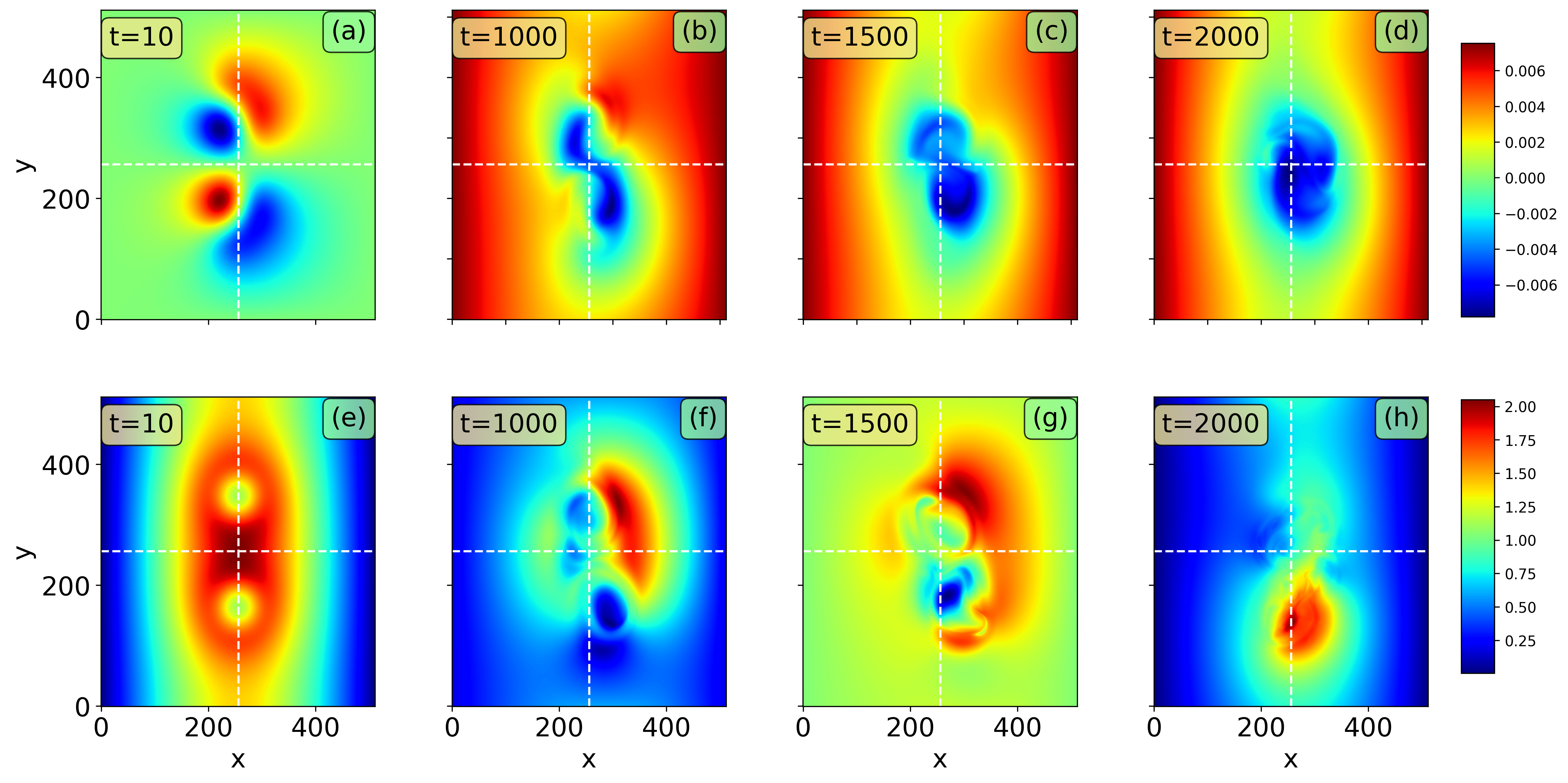}
    \caption{Time evolution of the electrostatic potential during filament merging for (a-d) the cold-ion case ($\tau=0$) and (e-h) the warm-ion case ($\tau=1.0$). Warm-ion effects generate asymmetric, dipolar potential structures that drive strong poloidal $E\times B$ flows and coherent rotation, in contrast to the predominantly radial flows in the cold-ion regime.}
    \label{fig:potential_merging}
\end{figure*}

The corresponding time evolution of the electrostatic potential is presented in Fig.~\ref{fig:potential_merging}. In the cold-ion regime (a-d), the $\phi$ remain largely symmetric throughout the interaction. The potential contours are predominantly aligned in the radial direction, producing a mainly radial $E\times B$ flow. As a consequence, the filaments experience a strong inward radial drift toward each other, consistent with efficient merging driven by direct current attraction and polarization-induced radial motion in the $y$- direction. The mono-polar potential is also shown in the merging time (Fig.~\ref{fig:potential_merging} h) as previously reported by Mondal et al. \cite{souvik_pop}. 

On the other hand, for the warm-ion regime (e-h), the $\phi$ structure is different because of the initial pressure term in the effective vorticity term as $\omega=\nabla_\perp^2\phi + \nabla_\perp^2p/n_i$.  As a results in the creation of asymmetric and dipolar potential structures, which produce significant poloidal $E\times B$ velocity. This behavior can be attributed to increased vorticity generation, $\omega=\nabla_\perp^2\phi + \nabla_\perp^2p/n_i$, induced by pressure gradients via the polarization response.
\begin{equation}
\partial_t \omega \sim \mathbf{b}\times\nabla p ,
\end{equation}
with $p=n(T_e+T_i)=nT_e(1+\tau)$. As $\tau$ grows, higher pressure gradients produce greater vorticity, resulting in persistent rotating motion. The transition of flow from radial to a combination of radial and poloidal motion substantially changes the interaction geometry of the filaments and is crucial in delaying the merging process.

\begin{figure}
    \centering
    \includegraphics[width=0.9\linewidth]{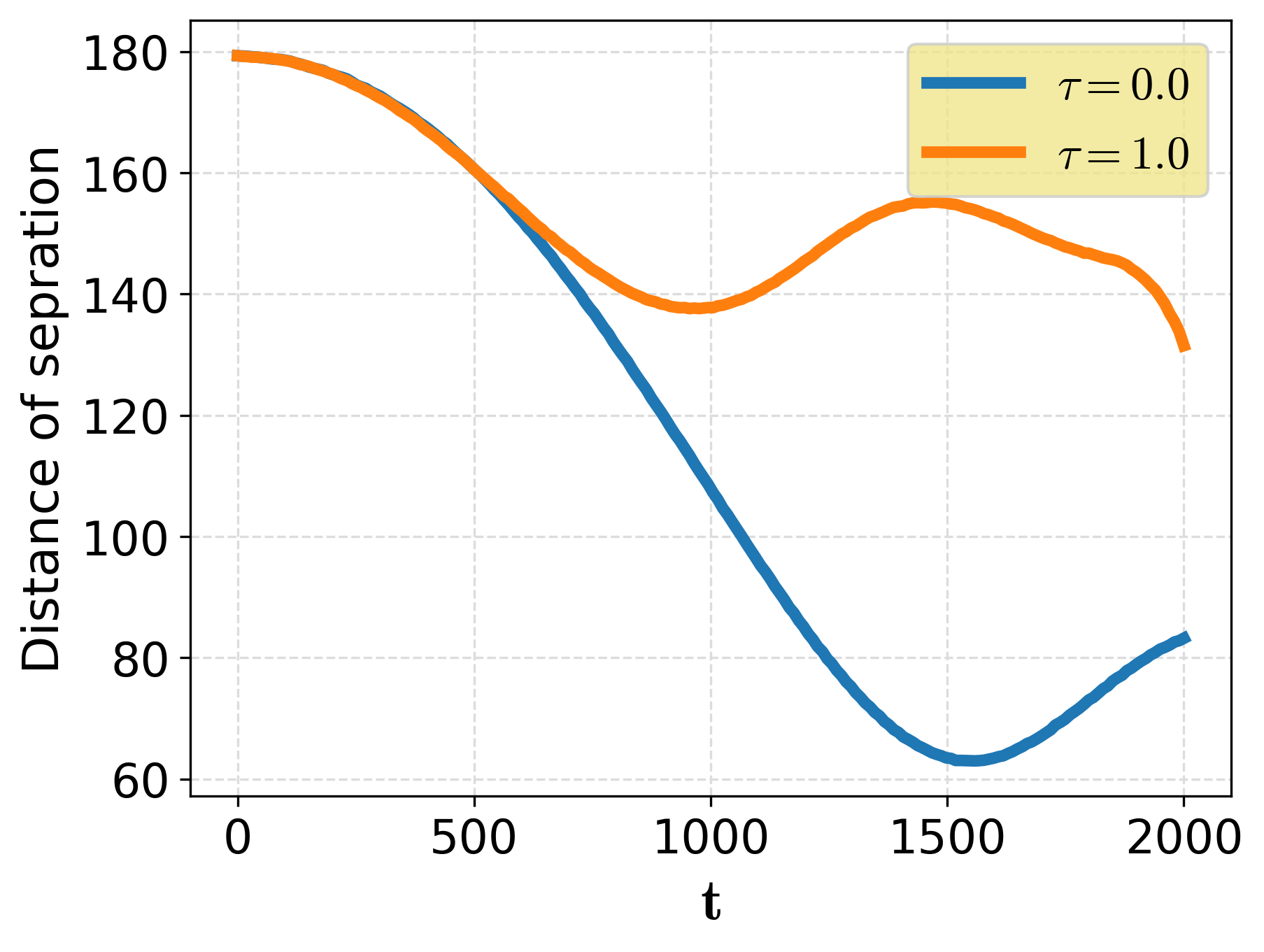}
    \caption{Time evolution of the separation distance between the centers of two unidirectional current-carrying filaments for the cold-ion ($\tau=0$) and warm-ion ($\tau=1.0$) cases. Finite ion temperature leads to a significantly slower reduction of separation, indicating delayed merging due to enhanced poloidal and rotational dynamics.}
    \label{fig:separation}
\end{figure}

The influence of altered density and potential evolution on filament interaction is measured by the temporal evolution of the separation distance between the filament, as seen in Fig.~\ref{fig:separation}. The distance of separation is calculated using the centre of mass (COM) position of the each filament. In the cold-ion case ($\tau=0$), the separation decreases rapidly and nearly monotonically, indicating efficient convergence and fast merging. This behavior reflects a predominantly translation-dominated interaction, in which the filaments respond efficiently to curvature-driven polarization and electromagnetic forces. On the other hand, the warm-ion case ($\tau=1.0$) exhibits a markedly slower reduction in the distance of the filament separation. There is a noticeable delay in the merging process when the separation steadily reduces over a long period of time.

\begin{figure*}
    \centering
    \includegraphics[width=0.95\linewidth]{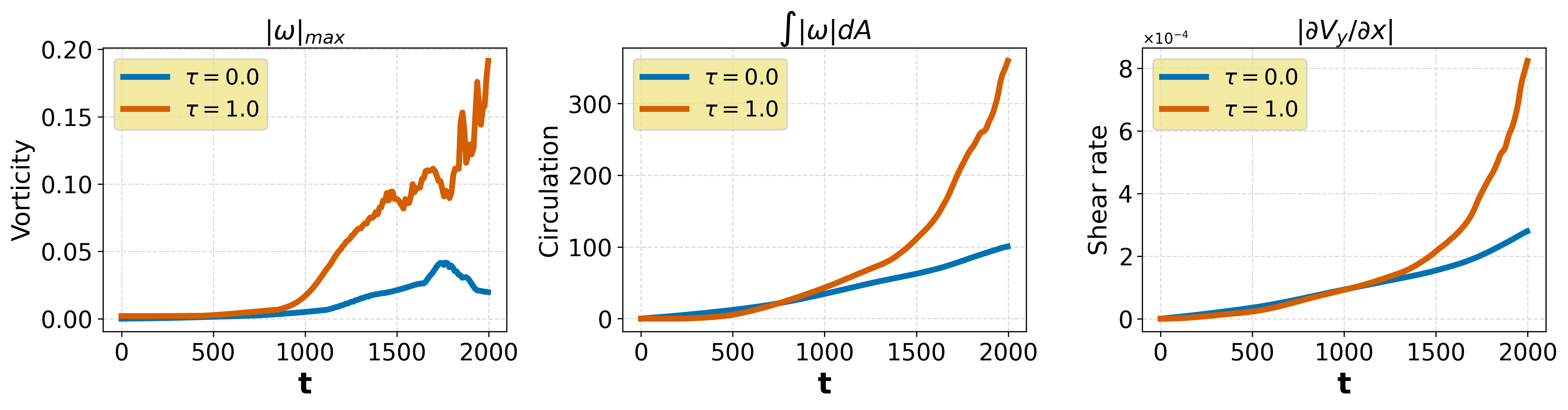}
    \caption{Time evolution of (left) peak vorticity $|\omega|_{\max}$, (middle) total circulation $\int |\omega|,dA$, and (right) shear rate $|\partial v_y/\partial x|$ for two interacting current-carrying filaments in the cold-ion ($\tau=0.0$) and warm-ion ($\tau=1.0$) regimes. The warm-ion case exhibits significantly enhanced vorticity generation, leading to a strong increase in circulation and shear at later times. This indicates a transition from translation-dominated dynamics in the cold-ion regime to rotation- and shear-dominated dynamics in the warm-ion regime, which redistributes kinetic energy into vortical motion and delays filament merging.}
    \label{fig:vort_diagnostic}
\end{figure*}

The delay in merging inside the warm-ion situation can be directly attributed to increased vorticity formation and the ensuing redistribution of flow energy into rotational and shear dynamics. Figure \ref{fig:vort_diagnostic} demonstrates that the maximum vorticity $|\omega|_{\max}$ and the total circulation $\int |\omega| \, dA$ significantly increases with ion temperature, indicating that a finite $\tau$ amplifies the pressure-driven polarization response. This corresponds with the generalized vorticity definition $\omega = \nabla_{\perp}^2 \phi + \nabla_\perp^2 p/n_i$ and its evolution $\partial_t \omega \sim \mathbf{b}\times\nabla p$, where $p = nT_e(1+\tau)$, suggesting that vorticity drive is proportional to $(1+\tau)$ \cite{Madsen2011_gyrofulid,Ricci2013_gyrokinetic}. The substantial increase in circulation suggests that a notable fraction of kinetic energy is transformed into coherent vortical motion, while the simultaneous increase in shear, quantified by $|\partial v_y/\partial x|$, denotes the development of intense differential $E\times B$ flows. The shear and rotational flows distort and transport the filaments azimuthally, hindering an effective center-of-mass approach. In contrast to the cold-ion regime, characterized by limited vorticity leading to predominantly radial $E\times B$ transport and rapid coalescence \cite{Garcia2006,dippolito_convective_2011}, the warm-ion regime is distinguished by rotation-dominated dynamics, wherein filaments undergo orbital motion and shear-induced deformation. Consequently, despite a stronger pressure gradient, the effective poloidal convergence decreases, resulting in a notable delay of filament merging.


\subsection{The influence of Ion Temperature on Single-Filament Dynamics}


In the previous section, we explained the impact of finite $\tau$ on filament interaction. To further understand the underlying physics, we now focus on the dynamics of a single isolated filament. This allows us to clearly identify how finite ion temperature modifies the density structure, electrostatic potential, flow patterns, and energy distribution without the complexity of multi-filament effects.

\begin{figure*}
    \centering
    \includegraphics[width=0.9\linewidth]{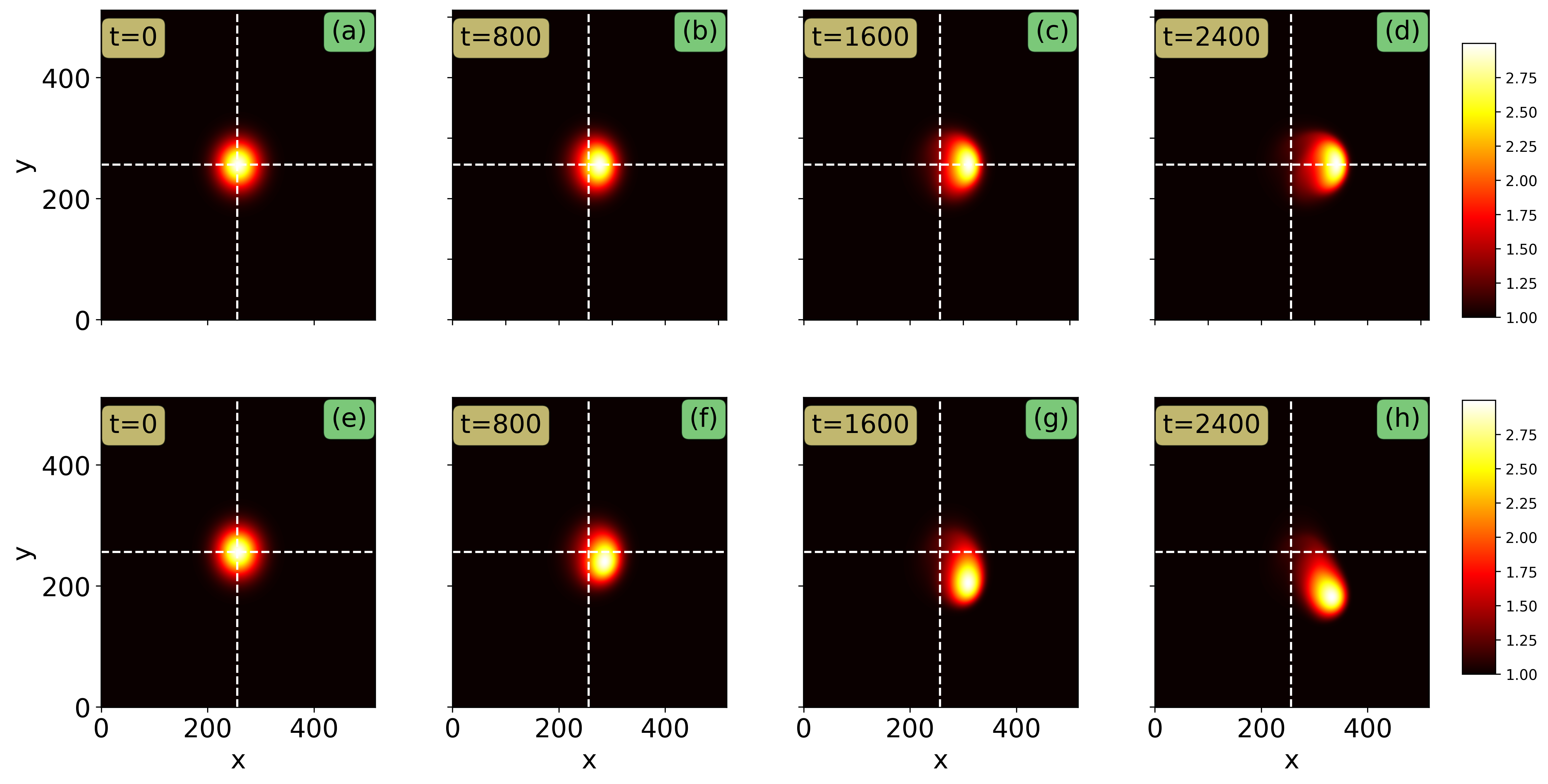}
    \caption{Time evolution of the density for an isolated filament in (a-d) the cold-ion case ($\tau=0$) and (e-h) the warm-ion case ($\tau=1.0$). In the cold-ion regime, the filament remains compact and approximately axisymmetric while propagating predominantly in the radial direction. In contrast, the warm-ion filament develops deformation, elongation, and tilting of the density contours, indicating the rotational dynamics driven by finite ion temperature effects.}
\label{fig:den_single}
\end{figure*} 

Figure~\ref{fig:den_single} shows the time evolution of the filament density for cold-ion ($\tau=0$) and warm-ion ($\tau=1.0$) cases. In the cold-ion regime (Fig.~\ref{fig:den_single}(a-d)), the filament retains an approximately axisymmetric Gaussian structure as it propagates radially. The density filament remains compact, and no significant deformation or rotation is observed, consistent with interchange-driven motion dominated by radial $E\times B$ advection. In contrast, the warm-ion filament (Fig.~\ref{fig:den_single}(e-h)) exhibits deformation at later times, including tilting and elongation in the poloidal direction. This distortion indicates the onset of rotational dynamics and reflects the enhanced coupling between pressure gradients and vorticity when ion temperature effects are included.

\begin{figure*}
    \centering
    \includegraphics[width=0.9\linewidth]{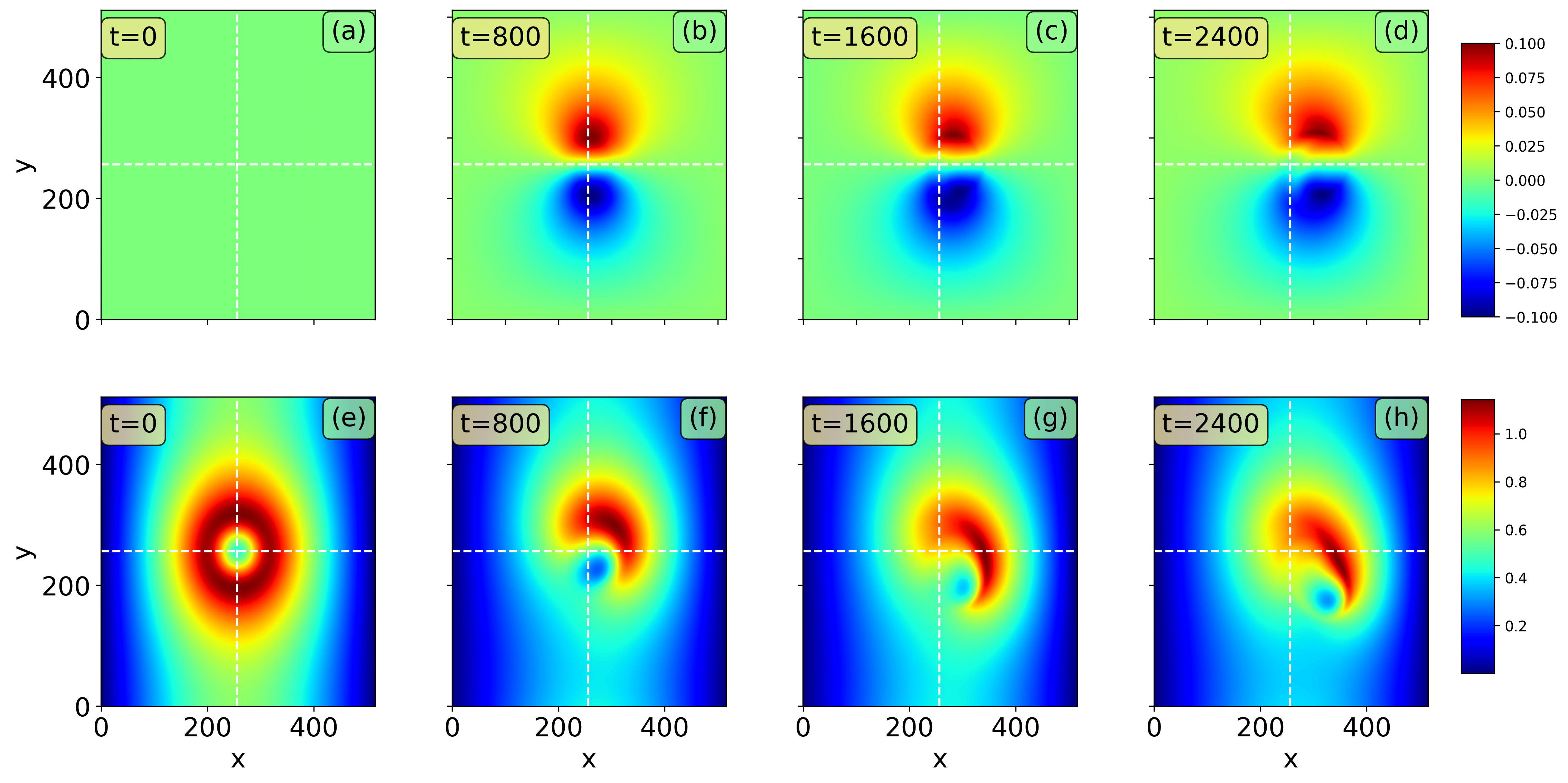}
    \caption{Time evolution of the electrostatic potential for an isolated filament in (a-d) the cold-ion case ($\tau=0$) and (e-h) the warm-ion case ($\tau=1.0$). The cold-ion filament develops a nearly symmetric dipolar potential structure that drives predominantly radial $\mathbf{E}\times\mathbf{B}$ motion. In contrast, the warm-ion filament exhibits strongly asymmetric and distorted potential contours, generating significant radial electric fields and sustained poloidal motion.}
\label{fig:phi_single}
\end{figure*}

The origin of this behavior is clarified by the corresponding electrostatic potential evolution shown in Fig.~\ref{fig:phi_single}. The potential satisfies the polarization response through the vorticity equation,
\begin{equation}
\omega = \nabla_\perp^2 \phi,
\end{equation}
with its temporal evolution governed by pressure-gradient-driven polarization currents,
\begin{equation}
\partial_t \omega \sim \mathbf{b}\times\nabla p,
\end{equation}
where $p = n(T_e + T_i) = nT_e(1+\tau)$. In the cold-ion case (Fig.~\ref{fig:phi_single}(a-d)), the potential retains a nearly dipolar structure aligned with the radial direction, producing an electric field that primarily drives outward motion. In contrast, the warm-ion case (Fig.~\ref{fig:phi_single}(e-h)) develops a strongly asymmetric and rotating potential pattern. This asymmetry directly generates a substantial radial electric field, breaking the radial symmetry of the filament motion and leading to significant radial as well as poloidal motion.

The resulting filament velocity is quantified using the density-weighted center-of-mass velocity,
\begin{equation}
\mathbf{V}_{\rm blob}(t) = \frac{\int \mathbf{v}_{E\times B} \, n \, dA}{\int n \, dA},
\qquad
\mathbf{v}_{E\times B} = \frac{\mathbf{b}\times\nabla\phi}{B}.
\end{equation}

\begin{figure*}
  \centering
  \includegraphics[width=0.45\linewidth]{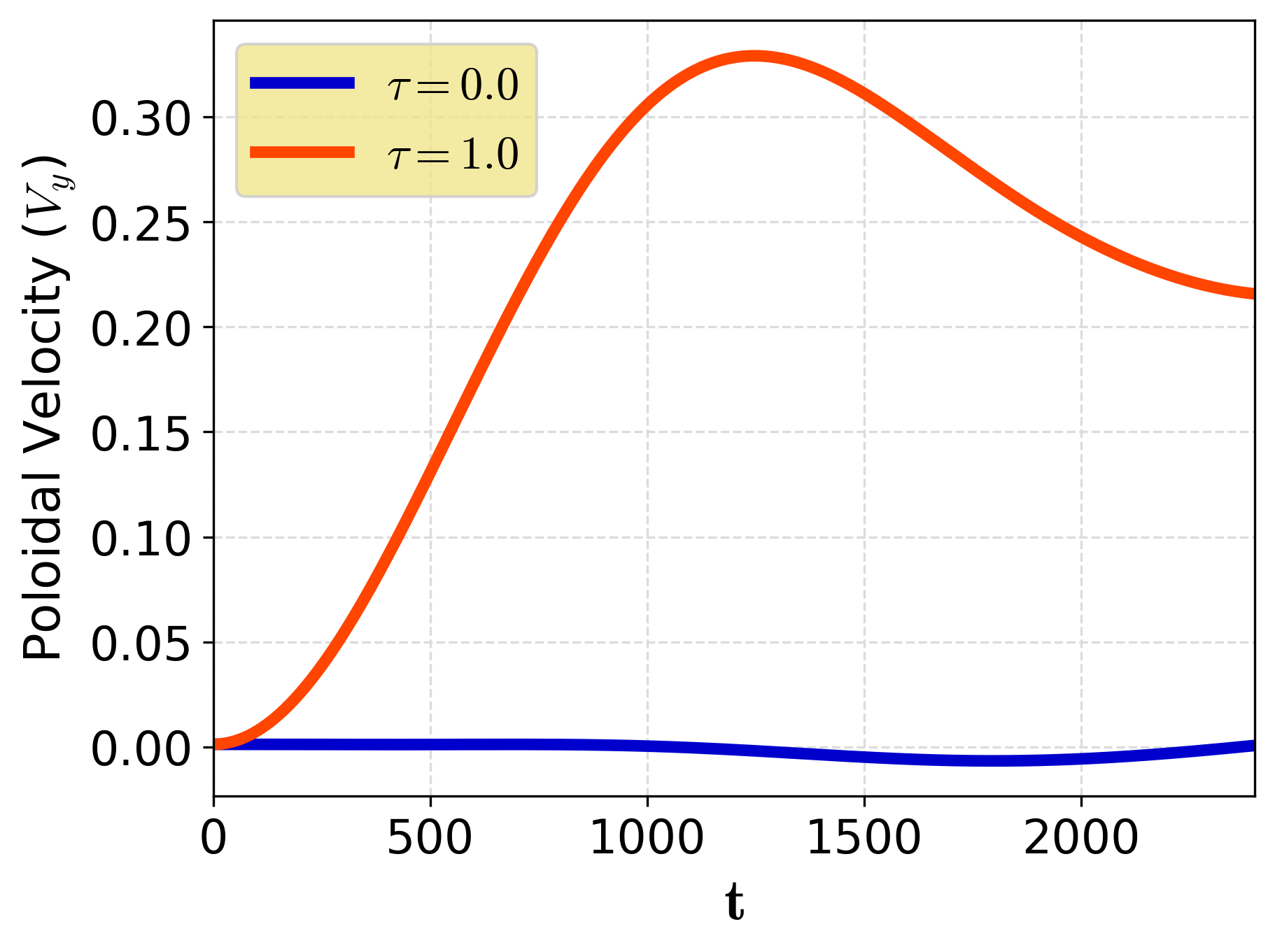}\hfill
  \includegraphics[width=0.45\linewidth]{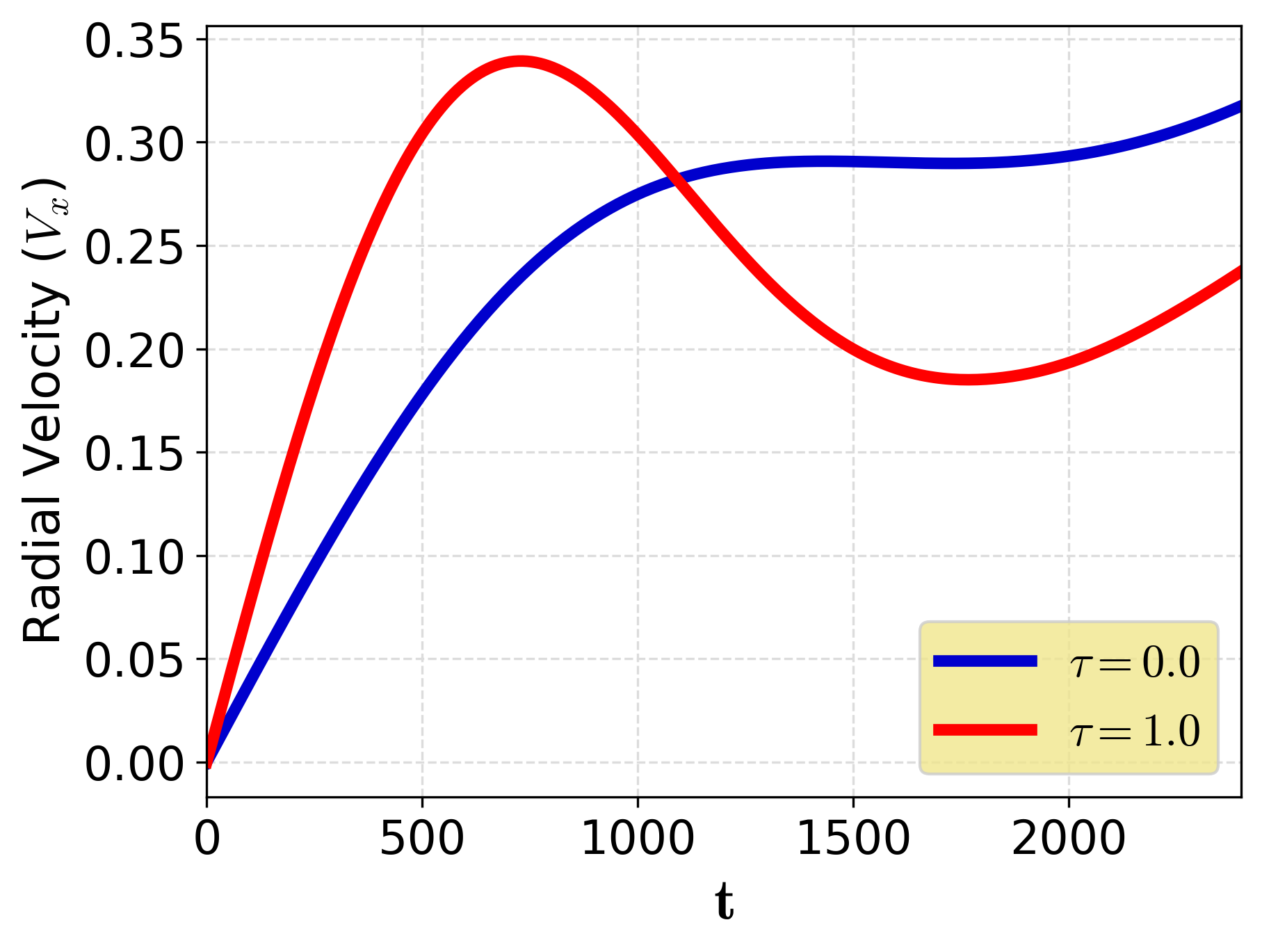} 
  \caption{Time evolution of the density-weighted center-of-mass (a) poloidal and (b) radial velocities of an isolated filament for cold-ion ($\tau=0$) and warm-ion ($\tau=1.0$) cases. The cold-ion filament propagates almost exclusively radially with negligible poloidal velocity. In contrast, the warm-ion filament develops a strong and sustained poloidal velocity, leading to curved trajectories and mixed radial-poloidal motion.}
\label{fig:velocity_single}
\end{figure*}

Figure~\ref{fig:velocity_single}(a) shows the poloidal velocity evolution. In the cold-ion case, the poloidal velocity remains negligible throughout the simulation, confirming that the filament dynamics are effectively one-dimensional. On the other hand, the warm-ion filament develops a strong poloidal velocity that becomes comparable to the radial velocity, which shows the emergence of rotational motion. The corresponding radial velocity evolution (Fig.~\ref{fig:velocity_single}(b)) shows that while warm-ion filaments initially accelerate more rapidly, their radial propagation is reduced at later times due to the redistribution of energy into poloidal motion.

\begin{figure*}
  \centering
  \includegraphics[width=0.45\linewidth]{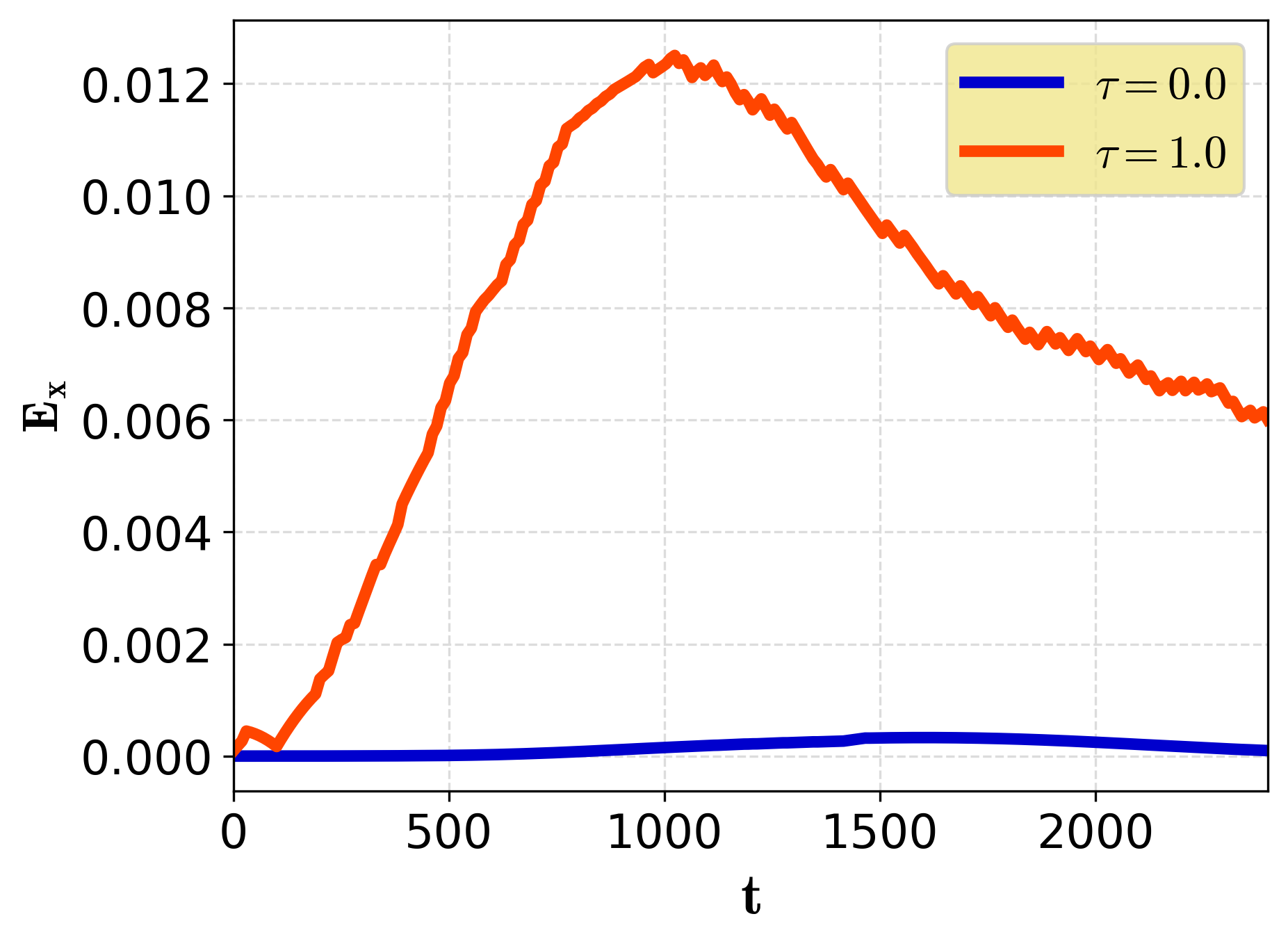}\hfill
  \includegraphics[width=0.45\linewidth]{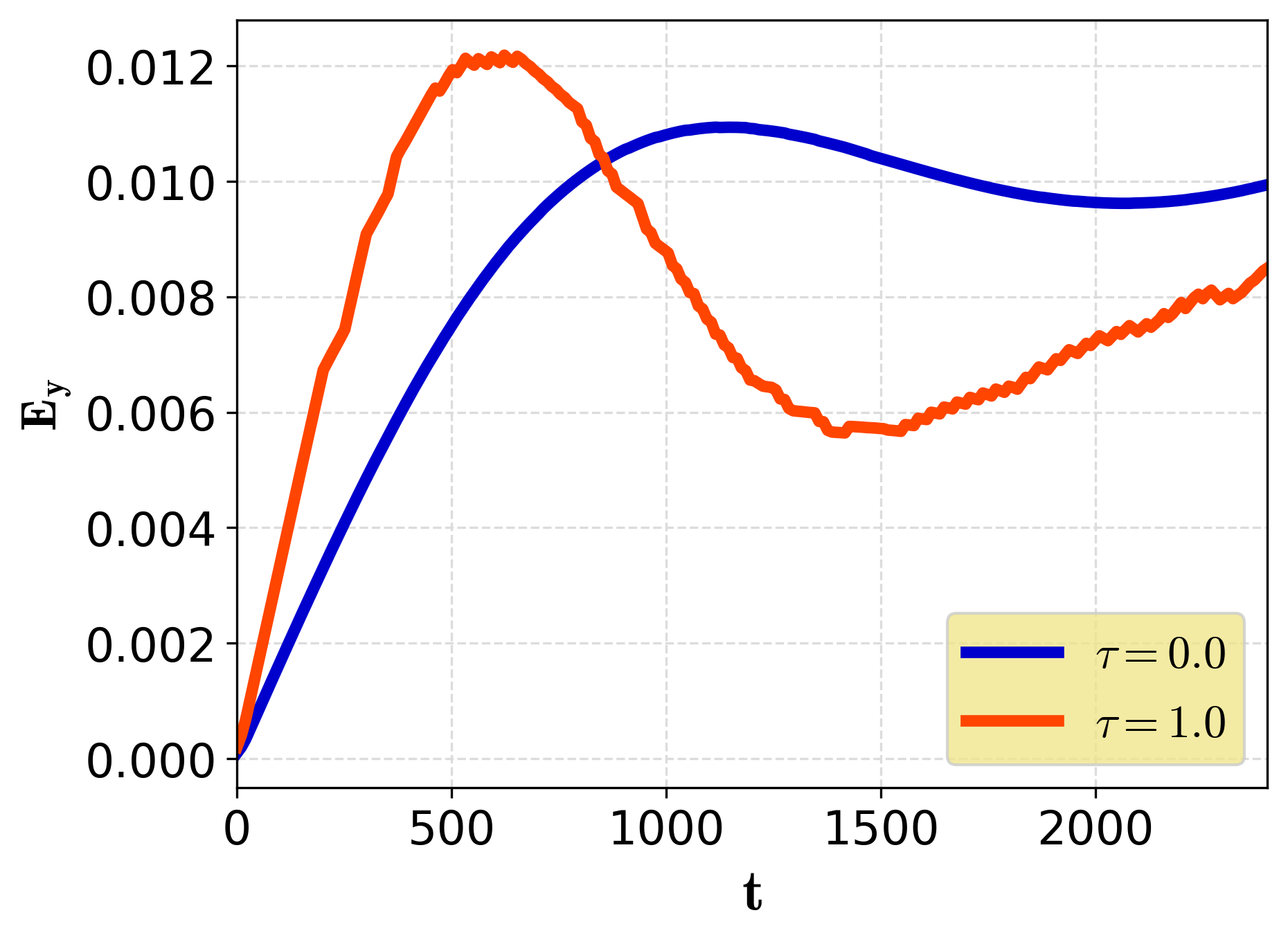} 
  \caption{Time evolution of the (a) radial electric field $E_x=-\partial_x\phi$ and (b) poloidal electric field $E_y=-\partial_y\phi$ for an isolated filament in cold-ion ($\tau=0$) and warm-ion ($\tau=1.0$) regimes. While the cold-ion case is dominated by the poloidal electric field, the warm-ion case exhibits a strong and persistent radial electric field, which drives substantial poloidal $E\times B$ motion and sustained rotational dynamics.}
\label{fig:efield_single}
\end{figure*}

This redistribution is driven by the electric-field structure shown in Fig.~\ref{fig:efield_single}, where the electric field components are computed as
\begin{equation}
E_x = -\partial_x \phi, \qquad E_y = -\partial_y \phi.
\end{equation}
In the cold-ion regime, the poloidal electric field ($E_y$) dominates and provides the primary interchange drive, while the radial electric field ($E_x$) remains weak. In contrast, the warm-ion case exhibits strong enhancement of both $E_x$ and $E_y$, with the poloidal electric field becoming comparable to or exceeding the radial component. This enhanced $E_x$ sustains poloidal $E_x\times B$ flows and promotes continuous vorticity generation, reinforcing the rotational filament dynamics.

\begin{figure*}
  \centering
  \includegraphics[width=0.45\linewidth]{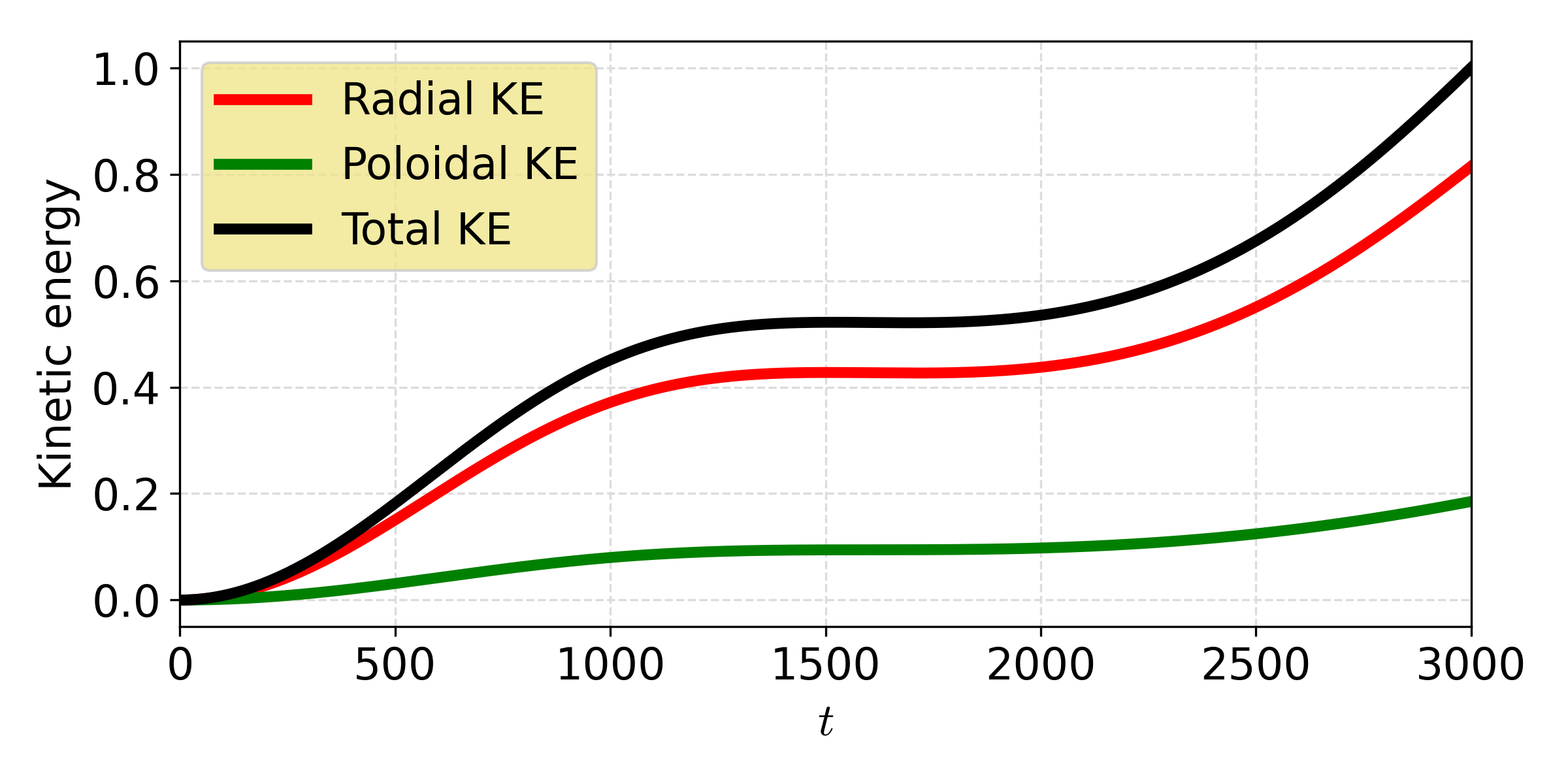}\hfill
  \includegraphics[width=0.45\linewidth]{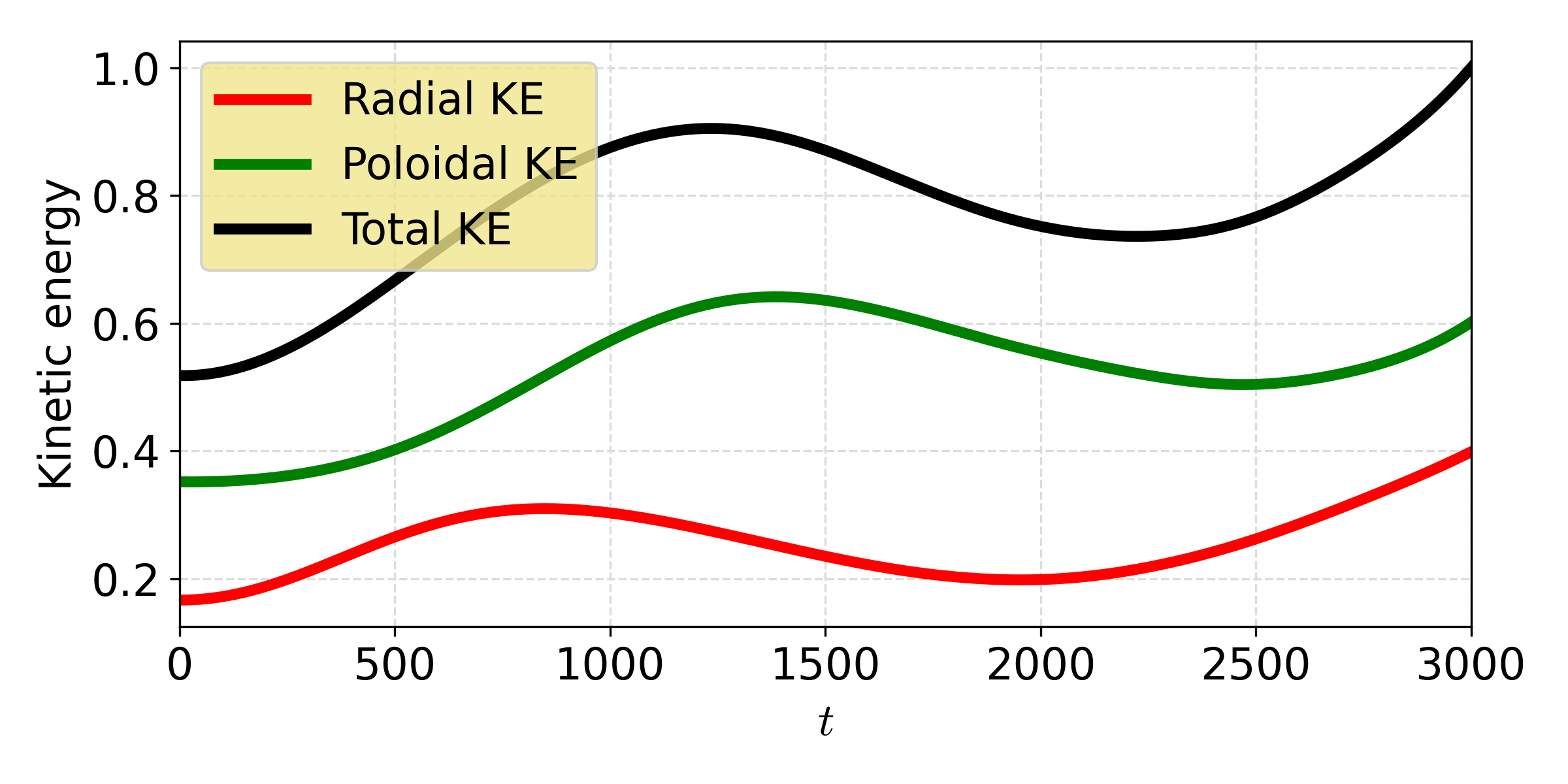} 
  \caption{Time evolution of the radial, poloidal, and total kinetic energy of an isolated filament for (a) the cold-ion case ($\tau=0$) and (b) the warm-ion case ($\tau=1.0$). In the cold-ion regime, the kinetic energy is dominated by the radial component, whereas in the warm-ion regime, a significant fraction of the kinetic energy is stored in poloidal motion, reflecting enhanced rotational dynamics driven by finite ion temperature effects.}
\label{fig:ke_single}
\end{figure*}

The energetic consequences of these modified flows are illustrated in Fig.~\ref{fig:ke_single}, which shows the decomposition of the kinetic energy,
\begin{equation}
E_K = \frac{1}{2}\int \left( v_x^2 + v_y^2 \right) dA,
\end{equation}
into radial, poloidal, and total components. In the cold-ion case (Fig.~\ref{fig:ke_single}(a)), the kinetic energy is almost entirely radial, confirming that the available free energy is efficiently converted into outward transport. On the other hand, the warm-ion case (Fig.~\ref{fig:ke_single}(b)) exhibits a substantially larger total kinetic energy, but with a significant fraction stored in poloidal motion rather than radial propagation.

\begin{figure*}
  \centering
  \includegraphics[width=0.45\linewidth]{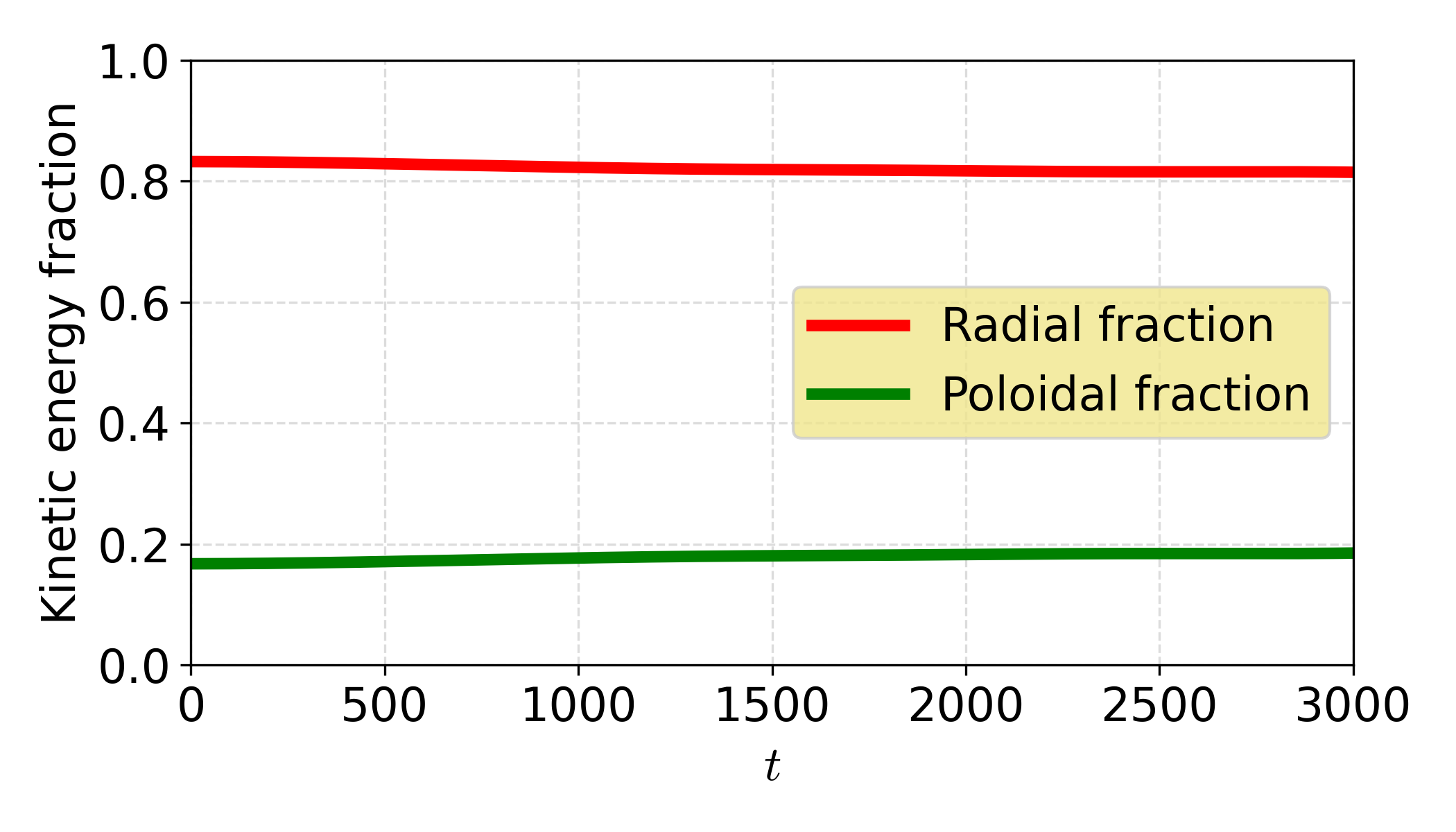}\hfill
  \includegraphics[width=0.45\linewidth]{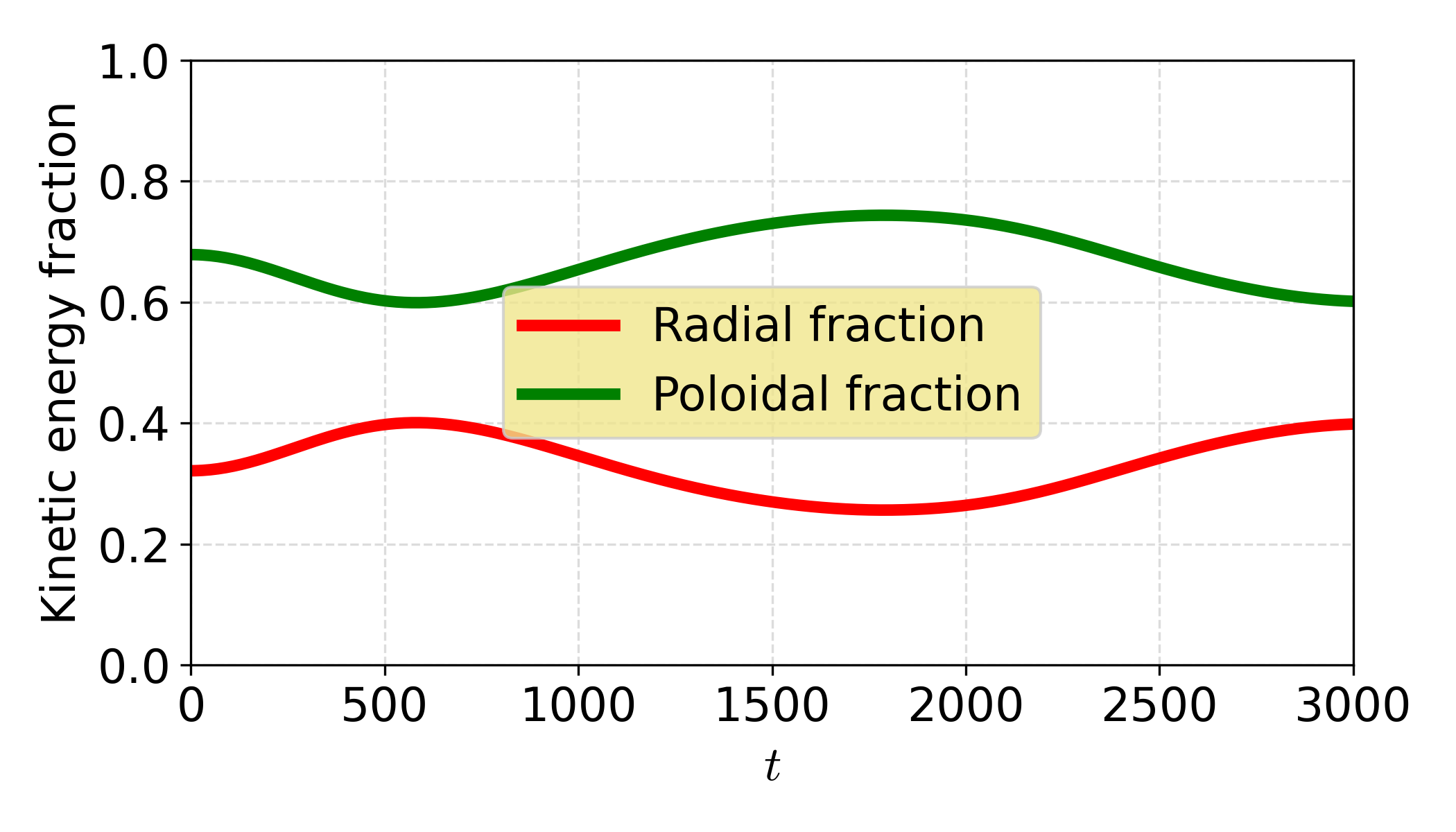} 
  \caption{Time evolution of the fractional contributions of radial and poloidal kinetic energy to the total kinetic energy for an isolated filament in (a) the cold-ion case ($\tau=0$) and (b) the warm-ion case ($\tau=1.0$). The cold-ion filament remains dominated by radial kinetic energy throughout the evolution, while the warm-ion filament exhibits a dominant and sustained poloidal kinetic energy fraction, indicating a fundamental redistribution of energy into rotational motion.}
\label{fig:ke_fraction}
\end{figure*}

This is further quantified by the kinetic energy fractions shown in Fig.~\ref{fig:ke_fraction},
\begin{equation}
f_r = \frac{E_{K,r}}{E_{K,\mathrm{tot}}}, \qquad
f_p = \frac{E_{K,p}}{E_{K,\mathrm{tot}}}.
\end{equation}
In the cold-ion regime, $f_r \approx 1$ throughout the evolution, reflecting purely radial dynamics. In contrast, the warm-ion regime exhibits a dominant poloidal kinetic energy fraction over extended periods, with oscillatory exchange between radial and poloidal components. This behavior reflects the competition between interchange acceleration and finite-ion-temperature driven rotational dynamics.

These results together show that a finite ion temperature fundamentally changes how single filaments behave. It does this by increasing the vorticity driven by pressure gradients, creating asymmetric potential structures, and shifting kinetic energy from moving outward to sustained rotation. Although filaments with warm ions have more energy, a significant amount of this energy is stored in the surrounding flows and vortices, rather than contributing to outward movement. This results in a significant change in the filament's behavior. \\

To understand how a finite ion temperature systematically changes filament dynamics, we perform a scan over $\tau$ and examine the resulting changes in filament velocity, the energy driving the filament, and how kinetic energy is distributed.

\begin{figure*}
  \centering
  \includegraphics[width=0.45\linewidth]{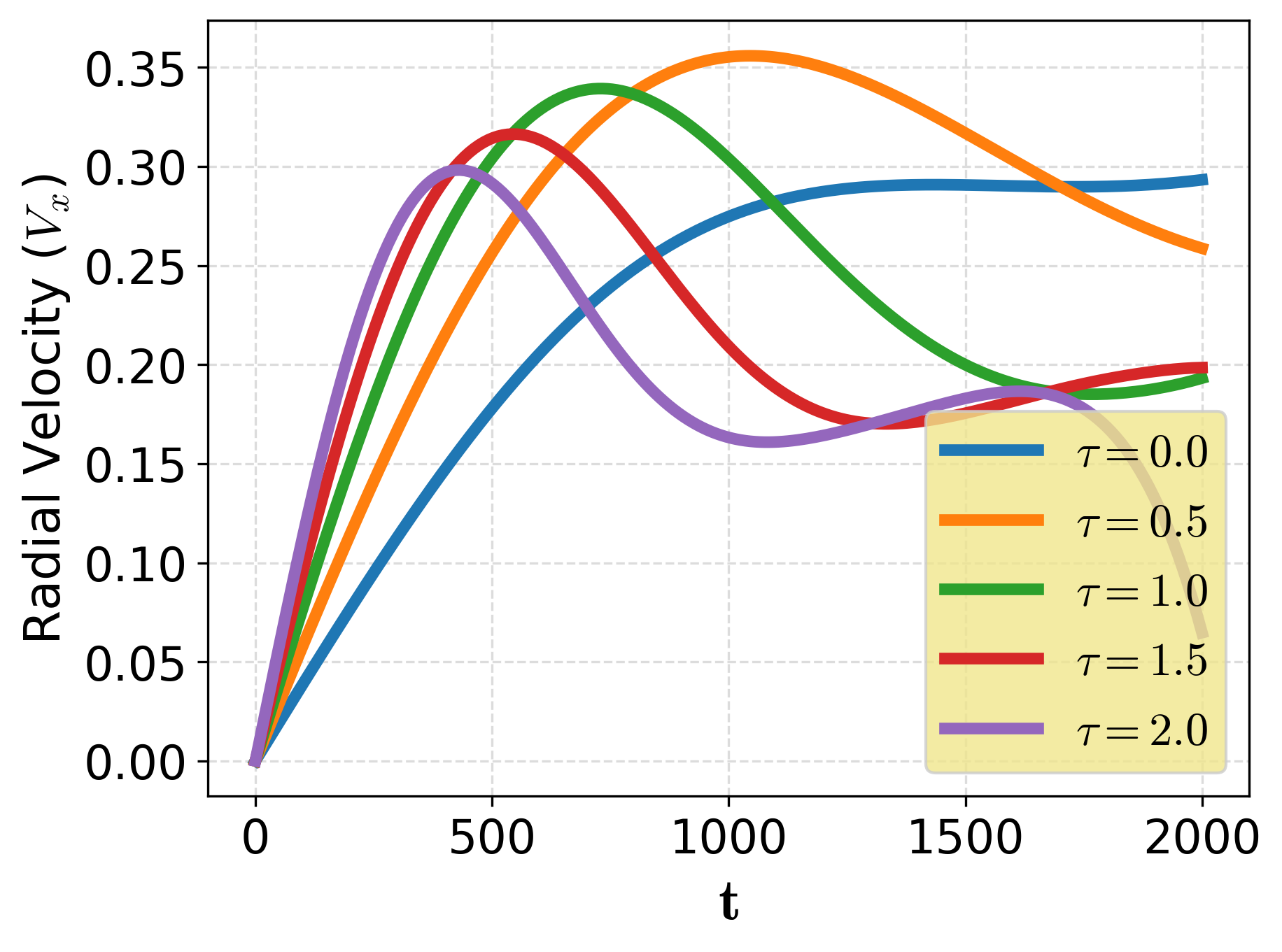}\hfill
  \includegraphics[width=0.45\linewidth]{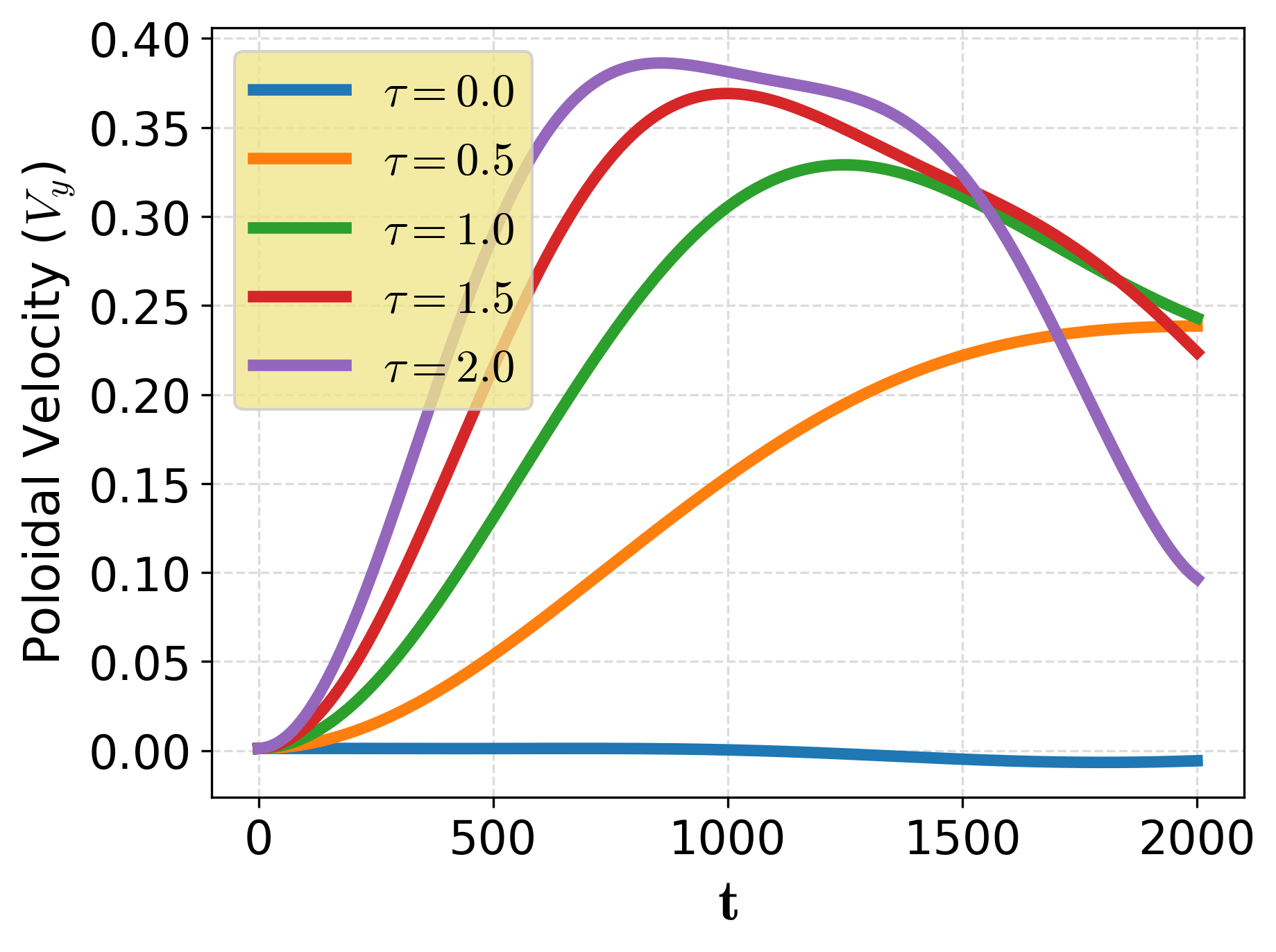}
  \caption{Time evolution of the density-weighted center-of-mass velocity of an isolated filament for different $\tau$: (a) radial velocity $V_x$ and (b) poloidal velocity $V_y$. Increasing $\tau$ enhances the overall filament motion and promotes a transition from radially dominated propagation to strong poloidal and rotational dynamics.}
\label{fig:tau_velocity}
\end{figure*}


Figure~\ref{fig:tau_velocity} shows the time evolution of the density-weighted center-of-mass velocities for different values of $\tau$. In the cold-ion limit ($\tau=0$), the filament accelerates radially and reaches a quasi-steady velocity, while the poloidal velocity remains negligible throughout the evolution. This behavior is characteristic of classical interchange-driven propagation dominated by radial motion. As $\tau$ increases, the early-time radial acceleration becomes stronger, reflecting the enhanced pressure-gradient drive. However, at higher $\tau$ the radial velocity exhibits a pronounced peak followed by a reduction at intermediate times, indicating that radial propagation is no longer the sole sink of injected free energy.

On the other hand, the poloidal velocity exhibits a strong and systematic increase with $\tau$. For $\tau\gtrsim1$, the poloidal velocity becomes comparable to or larger than the radial component and persists over long times. This demonstrates that finite ion temperature promotes sustained rotational motion by enhancing asymmetric potential structures and poloidal electric fields. As a result, filament trajectories become increasingly curved with increasing $\tau$.

\begin{figure*}
  \centering
  \includegraphics[width=0.9\linewidth]{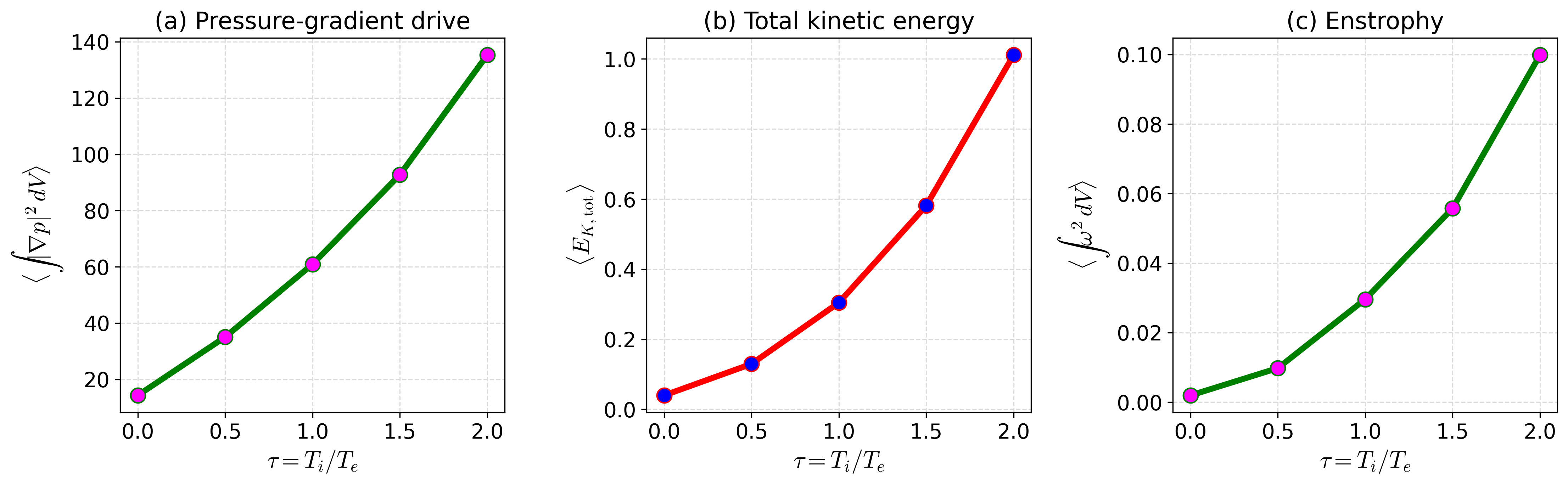}
  \caption{Dependence of key driving and energetic quantities on $\tau$: (a) volume-integrated pressure-gradient drive, (b) time-averaged total kinetic energy, and (c) time-averaged enstrophy. All three quantities increase monotonically with $\tau$, indicating enhanced free-energy injection and stronger dynamical activity at higher ion temperature. The rapid increase in enstrophy demonstrates preferential amplification of vorticity and rotational motion in the warm-ion regime.}
\label{fig:tau_drive_energy}
\end{figure*}

The energetic origin of this behavior is clarified in Fig.~\ref{fig:tau_drive_energy}, which shows the pressure-gradient drive, total kinetic energy, and enstrophy as functions of $\tau$. The pressure-gradient drive increases monotonically with $\tau$, consistent with the scaling
\[
p = nT_e(1+\tau), \qquad \nabla p \propto (1+\tau)\nabla n,
\]
indicating systematic injection of additional free energy at higher ion temperature. Correspondingly, the total kinetic energy increases rapidly with $\tau$, confirming that warm-ion filaments are energetically stronger. At the same time, the enstrophy, $\int \omega^2 dA$ with $\omega=\nabla_\perp^2\phi$, increases even more sharply. This indicates that finite ion temperature preferentially enhances vorticity generation and rotational activity rather than purely accelerating bulk radial flow.

 \begin{figure}
 \centering
  \includegraphics[width=0.9\linewidth]{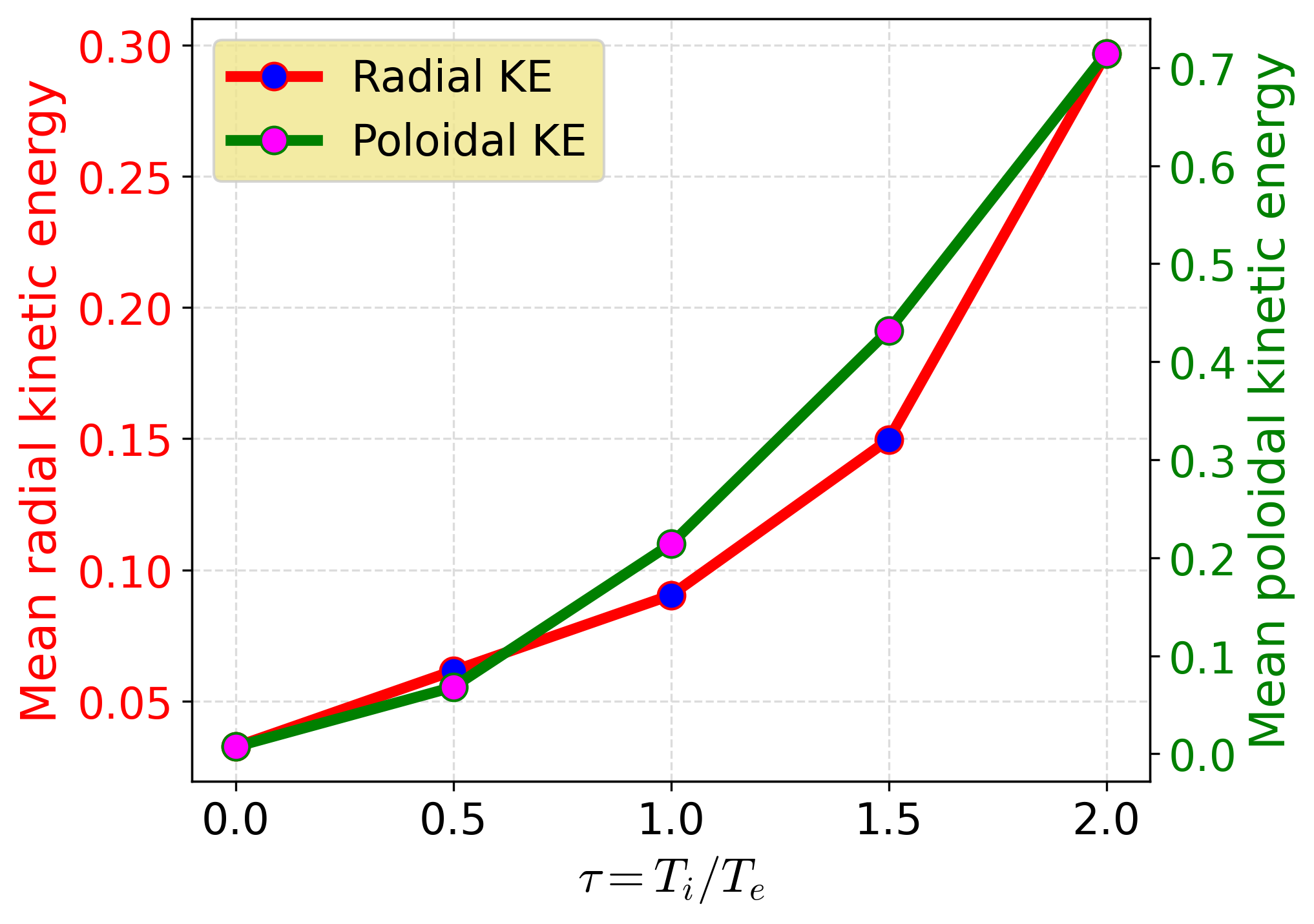}
  \caption{Time-averaged radial and poloidal kinetic energy components of an isolated filament as functions of $\tau$. The radial kinetic energy (left axis) increases moderately with $\tau$, while the poloidal kinetic energy (right axis) grows more rapidly and dominates at higher $\tau$. This demonstrates that finite ion temperature preferentially redirects injected free energy into poloidal and rotational motion rather than purely radial propagation.}
\label{fig:tau_ke_components}
\end{figure}

This redistribution of energy is quantified in Fig.~\ref{fig:tau_ke_components}, which shows the time-averaged radial and poloidal kinetic energy components as functions of $\tau$. While both components increase with $\tau$, the poloidal kinetic energy grows significantly faster and becomes dominant for $\tau\gtrsim1.5$. This finding demonstrates that the additional free energy introduced by finite ion temperature is primarily transformed into poloidal and rotational motion.

 \begin{figure}
 \centering
  \includegraphics[width=0.9\linewidth]{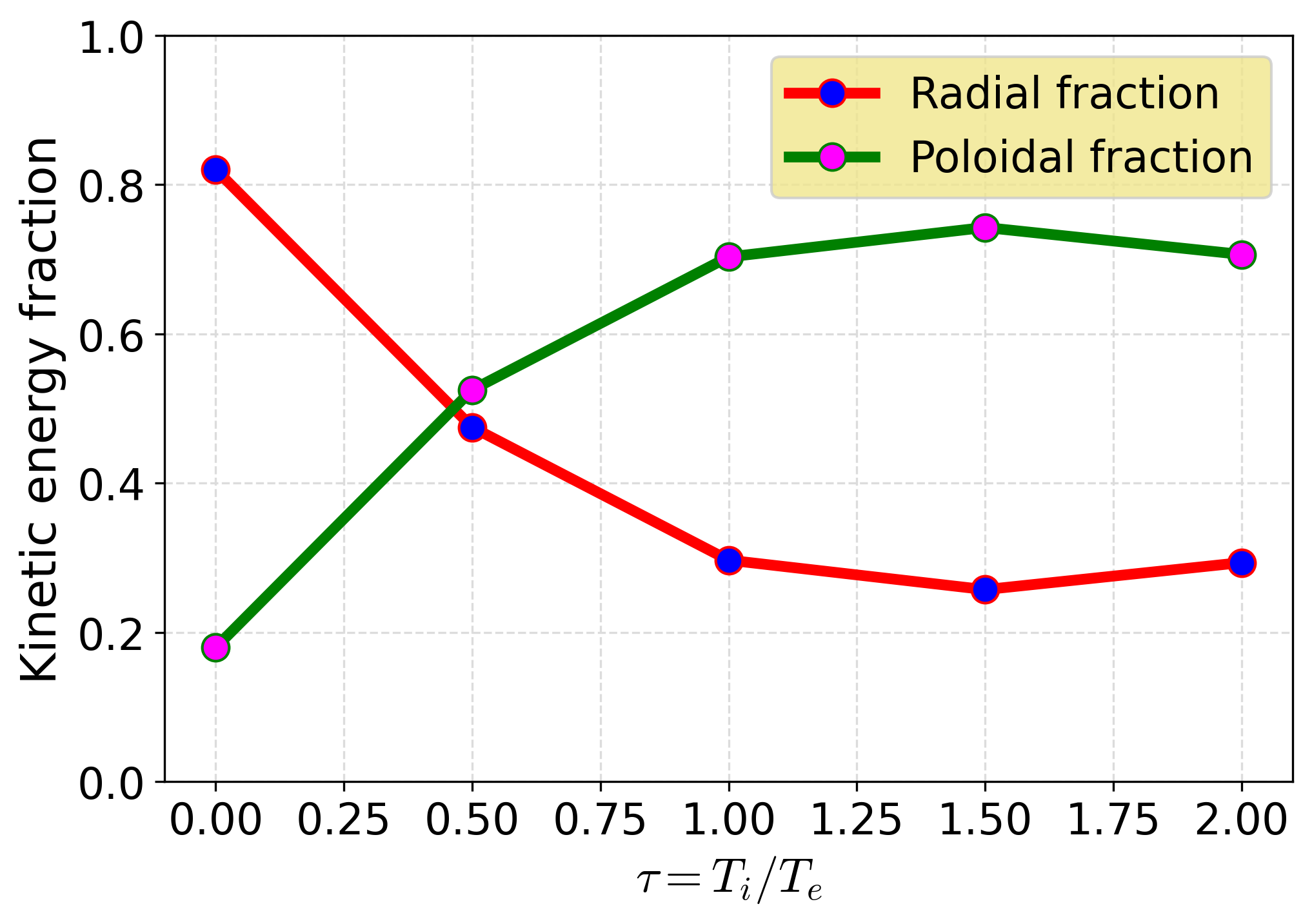}
  \caption{Time-averaged fractional contributions of radial and poloidal kinetic energy to the total kinetic energy as functions of $\tau$. In the cold-ion limit, the kinetic energy is dominated by the radial component. With increasing $\tau$, the radial fraction decreases while the poloidal fraction increases and becomes dominant, indicating a robust transition from radially dominated to rotation-dominated filament dynamics.}
\label{fig:tau_ke_fraction}
\end{figure}

Finally, Fig.~\ref{fig:tau_ke_fraction} presents the time-averaged kinetic energy fractions. The radial kinetic energy fraction decreases steadily with increasing $\tau$, whereas the poloidal fraction increases and reaches saturation at high $\tau$. This supports the conclusion that the transition from radially dominated to rotation-dominated dynamics is robust and persistent after time averaging. Although warm-ion filaments possess higher total kinetic energy, an increasing fraction of this energy is stored in poloidal motion rather than radial propagation. 

\section{Discussion and Conclusions}


This study examines the impact of finite ion temperature on the dynamics of ELM-like, current-carrying filaments using a 3D fluid model. By changing the ion-to-electron temperature ratio $\tau$, we demonstrate that ion temperature significantly influences both filament-filament interaction and the behaviour of isolated filament dynamics. The cold-ion scenario acts as a benchmark, and the results are systematically extended to warm-ion conditions.

We observe that finite ion temperature results in a delay in the merging process of interacting filaments. In the cold-ion regime, filaments retain compactness and predominantly traverse $y$- direction, and approach the fast merging. This behavior aligns with previous research on interchange-driven blob motion and the interaction of current-carrying filaments \cite{Garcia2006,dippolito_convective_2011,myra_current_carrying_filament_2007,souvik_pop}. Conversely, in the warm-ion scenario, filaments exhibit distortion and tilt and show a less efficient merging process.

The delay in merging is due to the strong rotating dynamics that appear at a finite ion
temperature. The warm-ion scenario shows a notable increase in vorticity, circulation, and shear,
which leads to strong poloidal flows. As a result, the filaments move slowly towards each
other; instead, they rotate, stretch, and move around each other. Previous studies on warm-ion and
gyrofluid models \cite{Madsen2011_gyrofulid,Wiesenberger2014,Held2016,Ricci2013_gyrokinetic} have also shown increases in rotational and vortical structures, suggesting that ion temperature is important for the generation of these dynamics.

We also studied the behavior of a single filament. In the cold-ion case, the filament stays mostly symmetrical and moves mainly outward. In contrast, the warm-ion filament shows asymmetry, tilting, and significant rotation, leading to curved paths and a mix of radial and poloidal motion. These observations agree with previous experiments and models, showing that higher ion temperatures reduce outward transport while increasing rotation in the scrape-off layerr \cite{Manz2013,Manz2015,Kocan2012}.

The primary outcome of this research is the recognition of energy redistribution as the main mechanism driving these alterations. Although an increase in ion temperature elevates the total available energy, a significant portion of this energy is diverted into poloidal and rotating motion rather than radial transport. This aligns with previous warm-ion scaling studies \cite{Manz2013,Manz2015}, but we demonstrate that this redistribution directly influences both filament propagation and interaction dynamics.

Our results indicate a clear transition from dynamics primarily dictated by radial motion in the cold-ion regime to those dominated by rotation at elevated ion temperatures. Consequently, this shift significantly impacts edge plasmas, particularly during edge-localized modes (ELMs), when ion and electron temperatures converge \cite{Kocan2012}. Consequently, to precisely describe filament behavior, transport phenomena, and structural changes, it is essential to account for warm-ion effects under these specific conditions.


\begin{acknowledgments}


The simulations were conducted on the Antya cluster at the Institute for Plasma Research (IPR). A. Sen expresses gratitude to the Indian National Science Academy (INSA) for the INSA Honorary Scientist designation.
 
\end{acknowledgments}

\bibliographystyle{unsrt}
\bibliography{citation}

\end{document}